\documentclass[12pt,preprint]{aastex}

\usepackage{url}

\shorttitle{Detecting variability in the NSVS}
\shortauthors{Shin et al.}
\slugcomment{Accepted to AJ}

\begin{document}

\title{Detecting Variability in Massive Astronomical Time-Series Data II: 
Variable Candidates in the Northern Sky Variability Survey}

\author{Min-Su~Shin}
\affil{Department of Astronomy, The University of Michigan, 500 Church Street, Ann Arbor, MI 48109, USA}
\email{msshin@umich.edu}

\author{Hahn~Yi}
\affil{Department of Astronomy, Yonsei University, Seoul, 120-749, Korea}
\email{yihahn@galaxy.yonsei.ac.kr}

\author{Dae-Won~Kim\altaffilmark{a}}
\affil{Department of Astronomy, Yonsei University, Seoul, 120-749, Korea}
\email{dakim@cfa.harvard.edu}

\author{Seo-Won~Chang}
\affil{Department of Astronomy, Yonsei University, Seoul, 120-749, Korea}
\email{seowony@galaxy.yonsei.ac.kr}

\and

\author{Yong-Ik~Byun}
\affil{Department of Astronomy and University Observatory, Yonsei University, Seoul, 120-749, Korea}
\email{ybyun@yonsei.ac.kr}

\altaffiltext{a}{Harvard Smithsonian Center for Astrophysics, Cambridge, MA 01238, USA}

\begin{abstract}
We present variability analysis of data from the Northern Sky Variability Survey (NSVS). 
Using the clustering method which defines variable candidates 
as outliers from large clusters, we cluster 16,189,040 light curves, having data points at more than 15 epochs, as variable and non-variable 
candidates in 638 NSVS fields. Variable candidates are selected depending on how strongly they are separated from the largest cluster and 
how rarely they are grouped together in eight dimensional space spanned by variability indices. All NSVS light curves are also 
cross-correlated to the {\it Infrared Astronomical Satellite}, AKARI, Two Micron All Sky Survey, Sloan Digital Sky Survey (SDSS), 
and {\it Galaxy Evolution Explorer} objects as well as known objects in the SIMBAD database. The variability 
analysis and cross-correlation results are provided in a public online database which can be used to 
select interesting objects for further investigation. Adopting conservative selection criteria for variable candidates, we 
find about 1.8 million light curves as possible variable candidates in the NSVS data, corresponding to about 10\% of our 
entire NSVS samples. Multi-wavelength colors 
help us find specific types of variability among the variable candidates. Moreover, we also use morphological classification from 
other surveys such as SDSS to suppress spurious cases caused by blending objects or extended sources due to the low angular resolution 
of the NSVS.
\end{abstract}


\section{Introduction}

Emerging projects in time-domain astronomy produce a large amount of time-series data, and allow discoveries 
of new variable sources and better understanding of known variability types \citep[see][for a review]{paczynski00,djorgovski01,bono03}. 
This new era needs a computationally intensive processing of massive time-series data with computational algorithms 
to recover interesting objects with  a broad range of variability types \citep{eyer06}.

Investigating time variability of astronomical objects begins with detecting any significant 
changes in brightness. Detection methods can be optimized to specific variability types. For example, 
image subtraction method is commonly used to detect supernovae, microlensing, and other variable sources 
\citep[e.g.,][]{alard98,wozniak00,gossl02,becker04,corwin06,yuan08}. Instead of detecting variability in 
images, recognizing variability in light curves can also be tailored to a particular type of 
variability signal such as transit \citep[e.g.,][]{protopapas05,renner08}.

We presented a new framework of detecting general variability types in massive time-series data by using a non-parametric 
infinite Gaussian mixture model (GMM) in our previous paper \citep[][hereafter, Paper I]{paper1}. In our approach, 
variable objects are considered as outliers from non-variable objects which should constitute a dominant fraction 
of given data, in multi-dimensional space spanned by several variability indices. In the results from the infinite GMM where 
each group is described by a multivariate Gaussian distribution \citep{robert96}, 
we recognize large groups as groups of non-variable objects, tagging outliers from these large 
groups as possible candidates of variable objects \citep[see][for other methods of clustering]{liao05}. 
The strength of our approach is based on the assumption that 
non-variable objects, which do not have enough signals of variable phenomena in their light curves, 
represent the dominant fraction of data and 
share the same systematic effects hidden in the given data such as sampling patterns and noise properties. 
Therefore, by extracting common properties of dominant non-variable objects 
from the given data, our approach can be less biased than choosing specific types of variability with 
assumptions of systematic patterns.

In this paper, the second of a series of papers, we apply our methodology to all data from 
the Northern Sky Variability Survey \citep[NSVS;][]{wozniak04a}. 
The NSVS catalog includes 
about 14 million\footnote{
Because some NSVS objects are included in multiple separate light curves from different 
observation fields, the total number of light curves is larger than that of objects.} 
objects with the optical magnitude ranging from 8 to 15.5 and declinations higher than $-38$ deg. 
These objects are so bright that deeper imaging surveys such as Pan-STARRS \citep{kaiser10} and 
Large Synoptic Survey Telescope \citep{tyson02} cannot produce useful photometric data 
due to saturation in their normal 
observation modes. Therefore, variability 
analysis of the current NSVS data will be still useful for investigation of bright stars even after deeper images are 
acquired in the future surveys. Moreover, the NSVS data have not been fully exploited because several previous 
trials have been focused on particular types of variable objects with specific criteria for variability signals 
\citep[e.g.,][]{wozniak04b,nicholson05,kinemuchi06,wils06,kelley07,kiss07,hoffman08,usatov08a,dimitrov09,schmidt09,hoffman09}.

Most NSVS objects are already included in other previous surveys. In particular, the NSVS catalog can cover most of the 
optical magnitude range of objects detected in {\it IRAS} \citep{clegg80}, 
Two Micron All Sky Survey \citep[2MASS;][]{kleinmann92}, and {\it Galaxy Evolution Explorer} \citep[GALEX;][]{bianchi99} observations. 
Since colors can be used to identify the particular types of spectral energy distributions, associating variability analysis 
to colors can be an effective way to discover new variable sources at a specific phase of stellar evolution such as Mira-type variables 
\citep[e.g.,][]{pojmanski05} or extragalactic variable objects such as quasars \citep[e.g.,][]{bianchi07}. 
Associating the NSVS objects with other optical surveys is also 
critically important. Since the NSVS data do not have a high angular resolution with small telescopes, 
a large fraction of objects might be affected by blending, poor positioning, and incorrect identification of 
extended objects as stars in the NSVS data. 
Our new variability analysis of the whole NSVS data will help others find specific variable sources by combining their data and 
these other multi-wavelength surveys.

This paper is organized as follows. 
In Section \ref{sec:gmm}, we explain the application of the infinite GMM to the NSVS data. 
In Section \ref{sec:sep}, we describe how to extract variable candidates by using the results from the GMM. 
In Section \ref{sec:web}, we explain the Web database of our variability analysis. 
Properties of variable candidates are explained in Section \ref{sec:var}, considering their associations 
to archival data. Finally, summary and discussions are given in the last section.

\section{Method}\label{sec:gmm}

\subsection{Data and variability indices}

We extract 16,189,040 light curves from the NSVS data \citep{wozniak04a} by limiting samples that have more than 
15 good photometric data points. The good photometric points are defined as not having bad photometry flags: SATURATED, 
NOCORR, LONPTS, HISCAT, HICORR, HISIGCORR, and RADECFLIP \citep[see][for meanings of the flags]{wozniak04a}. Since one object 
can be included as several light curves in different observation fields in the NSVS catalog, the number of objects is 
slightly smaller than that of the light curves in our sample. Here, we consider each light curve as a separate entity. 
Among 644 observation fields of the NSVS data, 
six fields (123c, 145c, 146c, 147c, 156c, 156d) do not have any light curves having more than 15 good photometric data points.

In addition to six variability indices used in Paper I, we also derive skewness and kurtosis from each light curve. 
The definitions of these indices are summarized in Table \ref{tab:var}. 
$\sigma/\mu$, $\gamma_{1}$, and $\gamma_{2}$ are not sensitive to structure of light curves. However, estimating them is 
computationally cheap, describing simple low-order patterns of light curves easily. 
We note that our definitions of skewness and kurtosis is not the same as the 
traditional measures \citep[see][for a discussion]{joanes98}. 
Other five indices describe more complex patterns in light curves. $Con$ represents how many sets of three consecutive data points are at least 
2 $\sigma$ fainter or brighter than the median magnitude, tracing continuous variations in light curves \citep{wozniak00}. 
$\eta$ measures the ratio of the mean square successive difference to the sample variance \citep{vonneumann41}. 
$J$ and $K$ have been commonly used for multi-band light curves although these can be estimated for single-band 
light curves \citep{stetson96}. Here, we use only 
single-band light curves with a slightly modified definition which uses sequential pairs of data points. Finally, we also use the analysis of variance (ANOVA) statistic which is useful for identifying 
periodic signals \citep{schwarzenberg96,shin04}. The maximum value of the ANOVA (AoVM) is used to measure the strength of periodicity. 
Even with an incorrect period of the light curve, AoVM can be a valuable quantity that infers periodicity \citep{shin07}.

\subsection{GMM and results}

We follow the same procedure of the infinite GMM as in our Paper I. The GMM is derived for each NSVS observation field with 
the eight variability indices. Even though the GMM converges much earlier than 100 iterations, we conduct 100 iterations 
as shown in Paper I. Since finding 
large groups is our main concern to find out non-variable objects from the samples, 
the number of iterations is not an important factor which affects how many small groups can be recovered. 
Each GMM component\footnote{
We use {\it component} as the same term as {\it cluster} and {\it group} in this paper. {\it Component} is a specific 
term for the GMM.} 
is described by a multivariate Gaussian distribution with its mean, i.e. center, and covariance matrix:
\begin{equation}
p_{m}({\bf x}) = \frac{1}{(2\pi)^{D/2}\vert\Sigma_{m}\vert^{1/2}} 
\exp(-\frac{1}{2}({\bf x} - {\bf \mu_{m}})^{T}\Sigma_{m}^{-1}({\bf x} - {\bf \mu_{m}})),
\label{eq:gaussian}
\end{equation}
where $m$ is an index of a mixture component, ${\bf x} = (\sigma/\mu, \gamma_{1}, \gamma_{2}, {\rm Con}, \eta, J, K, {\rm AoVM})$ 
is a vector of parameters, and $D$ is the number of parameters 
(in our case $D = 8$). Furthermore, ${\bf \mu_{m}}$ is a vector of mean 
values (i.e., mixture centers), and $\Sigma_{m}$ is the covariance matrix of 
the Gaussian distribution.

\begin{figure}
\plottwo{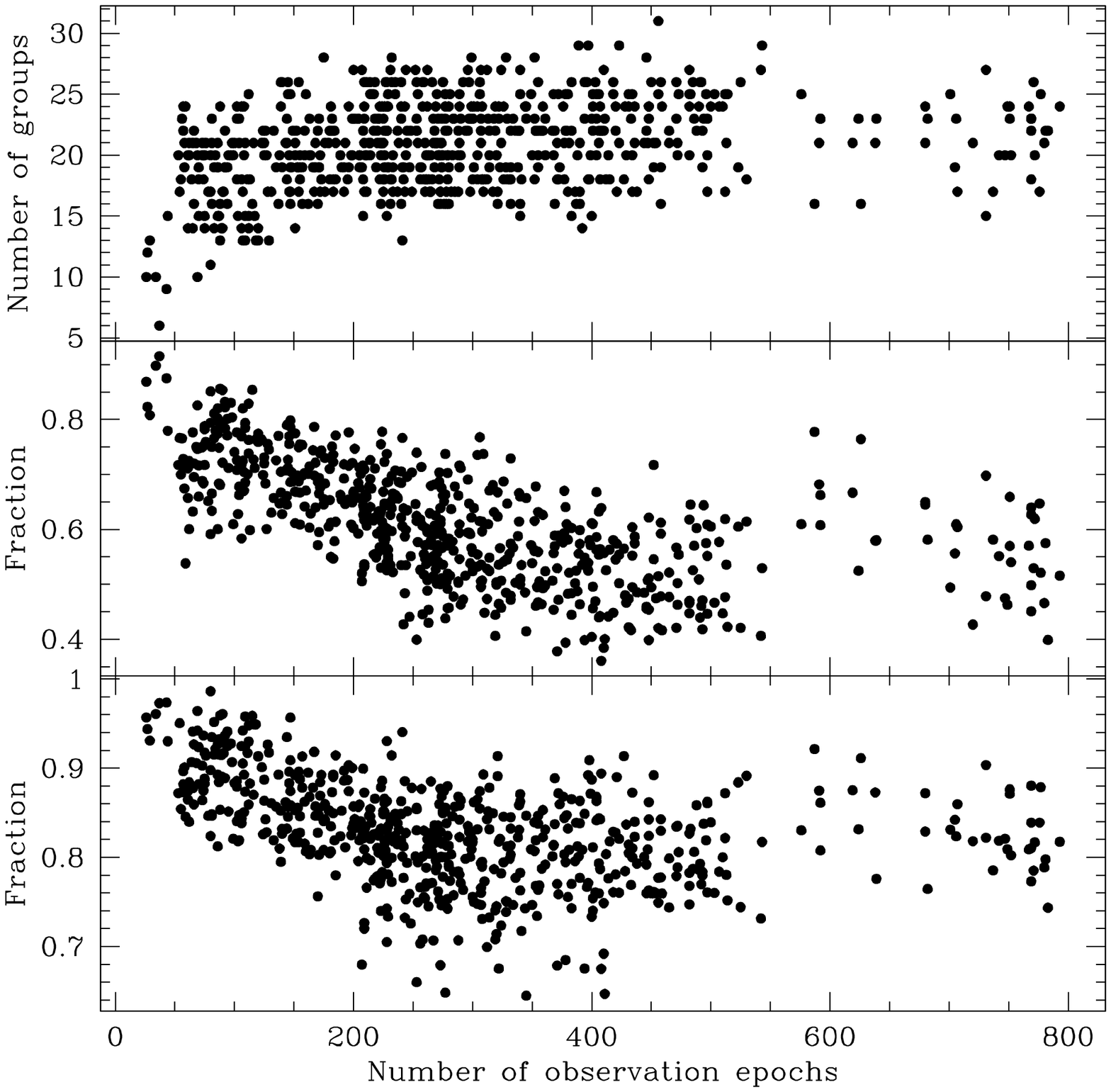}{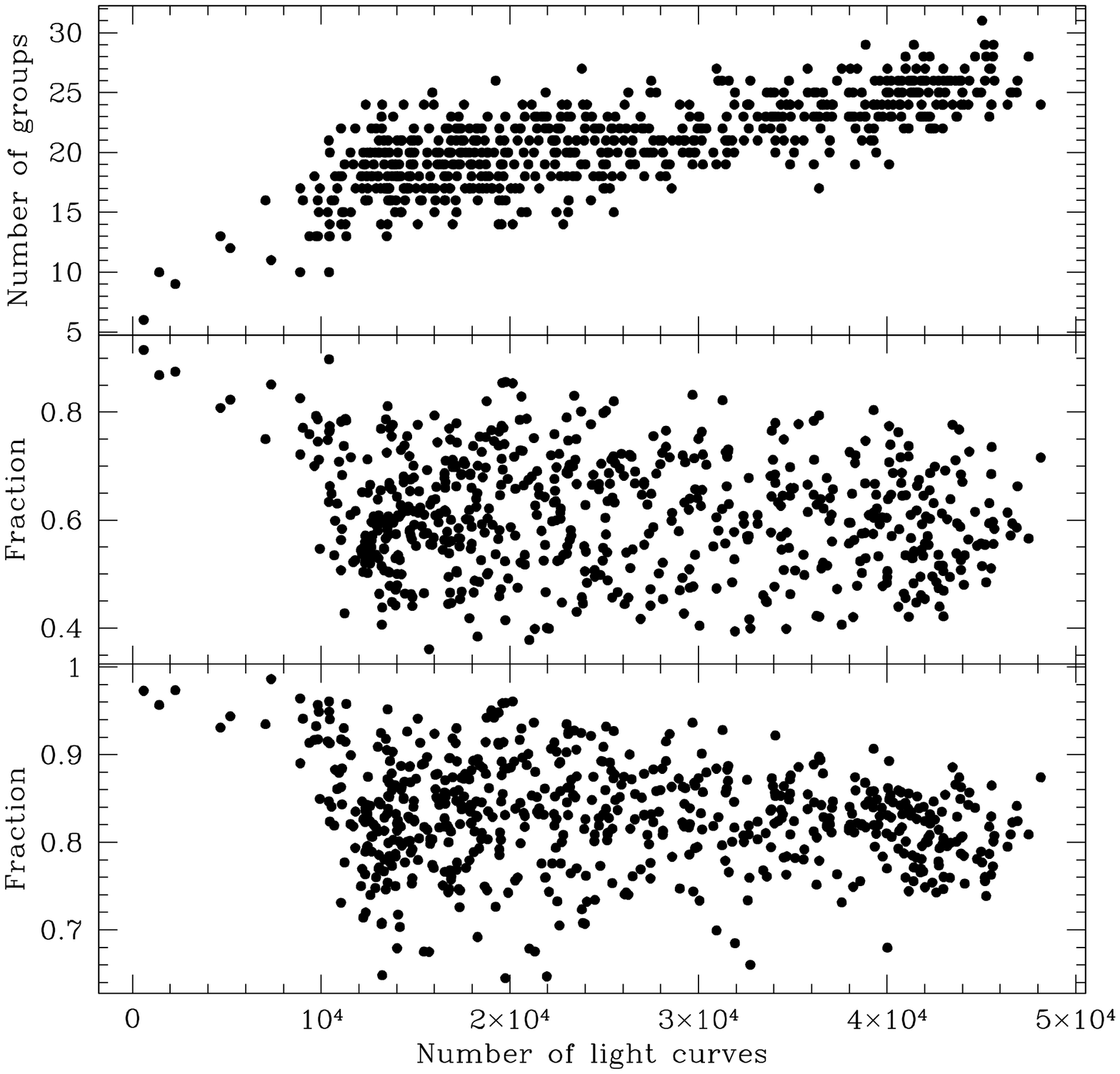}
\caption{GMM results with respect to the number of observation epochs (left) and the number of light curves (right). 
From top to bottom, each panel shows the number of groups, the number fraction of the largest group, and the number fraction of the top 
three large groups, respectively.}
\label{fig:gmm}
\end{figure}

Figure \ref{fig:gmm} shows how many groups are recovered with the infinite GMM and how many light curves are included in 
large groups. Since a small number of observation epochs provide poor sampling of light curves and have a low probability 
of detecting variability, fields having few observation epochs (i.e., few observed frames) are expected 
to have a small number of groups, while large groups dominate 
the whole population of light curves. In fields with a large number of light curves, the infinite GMM recovers many groups 
because these fields are likely to include various kinds of variable light curves. But the fraction of data included in 
the large groups does not change simply as a function of the number of light curves because 
the total number of light curves in the field does not affect the probability that a single light curve represents a 
non-variable object.

We measure the Davies -- Bouldin (DB) index in each field to check systematic differences of the GMM results. The DB index 
is commonly used to measure the compactness of clusters and separations among them \citep{davies79,vendramin10}. The index is defined as 
\begin{equation}
{\rm DB} ~=~ \frac{1}{n_{c}} \sum_{i=1}^{n_{c}} \max_{j=1,\dots,n_{c},~i \neq j} \left( \frac{s_{i} + s_{j}}{d_{ij}} \right),
\label{eq:DB}
\end{equation}
where $n_{c}$ is the number of clusters, $s_{i}$ is the average distance of data points included in the cluster $i$ with respect to its center, and 
$d_{ij}$ is the distance between centers of two clusters $i$ and $j$. Distance is defined as the $L^{2}$ norm which is one of the simple 
measurements \citep[see][for a discussion]{yu06}. As each cluster is compact and well separated from others, 
this DB index decreases.

\begin{figure}
\plotone{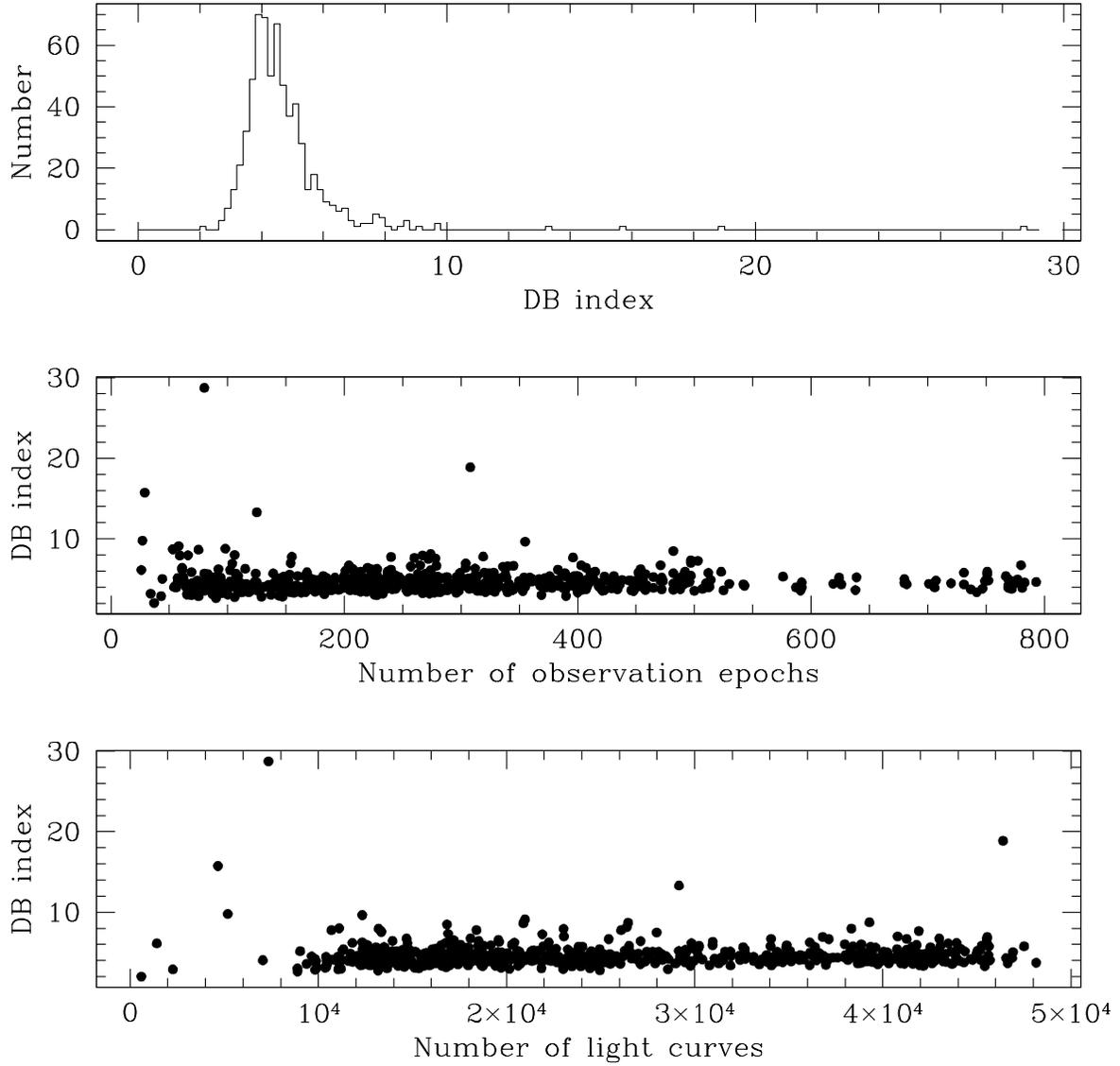}
\caption{Distributions of the DB index. When the DB index is low, groups are compact and well separated from others. 
The NSVS field 100c shows the highest DB index $\sim$ 29.}
\label{fig:DB}
\end{figure}

Strong systematic differences of the DB index do not appear among the 638 NSVS fields as shown in Figure \ref{fig:DB}. In most fields, 
the groups in the GMM results have a DB index smaller than 10. We do not find any strong systematic dependence of the DB index on the number of 
frames and the number of light curves in each field. The highest DB index $\sim$ 29 is found in the field 100c which has a small number 
of frames and light curves. Moreover, in that field the number fraction of the top three groups is highest among the sample fields.

\subsection{Largest cluster}

\begin{figure}
\plotone{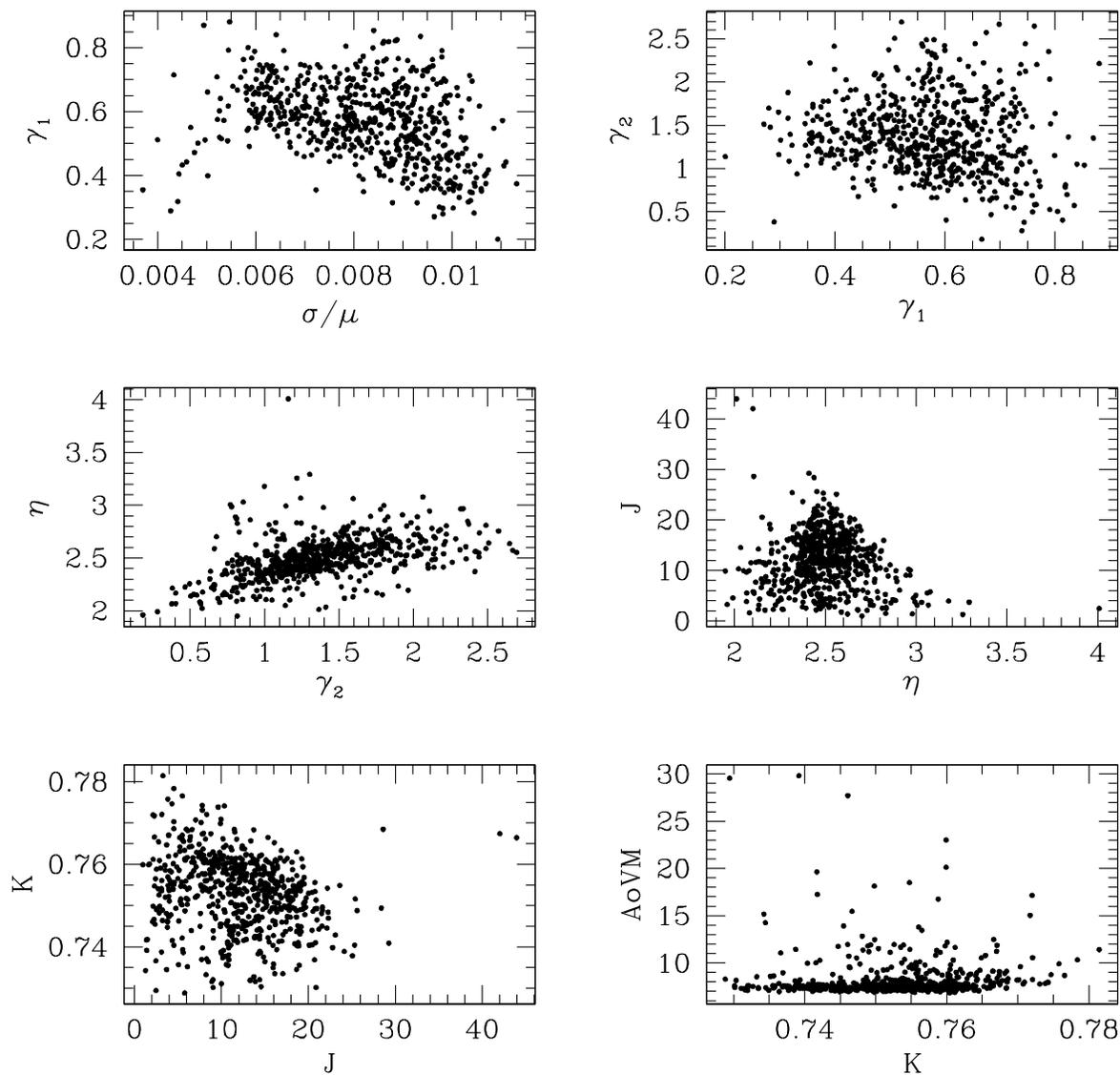}
\caption{Center coordinates of the largest cluster. Since $Con$ is 1 in all fields, we do not show its distribution here. 
The distribution shows that there are field-by-field variations of systematic effects which produce 
variations of the largest cluster's central position in the eight-dimensional space.}
\label{fig:center}
\end{figure}

Center coordinates of the largest cluster show a significant variation for different fields. Since each NSVS field 
has different characteristics, it is not surprising to see this variation shown in Figure \ref{fig:center}. Importantly, 
$\gamma_{1}$ and $\gamma_{2}$ are not close to zero in contrast to the expectation from assuming a normal distribution for 
light curves. These non-zero $\gamma_{1}$ and $\gamma_{2}$ imply that a dominant fraction of light curves are not well 
described by a normal distribution due to sampling or systematic observational effects. The same implication is also found 
by $\eta$ which has an expected average $\sim$ 2 for a normal distribution \citep{williams41}. Figure \ref{fig:center} proves the 
importance of including several variability indices which catch different features of light curves with different sensitivities as 
we showed in Paper I.

We also examine covariance matrices of the largest cluster in each field. Comparing the absolute values of the covariance 
matrix elements, the variances of $J$ or AoVM have maximum values in the largest cluster of 
all 638 fields. In 604 fields, the variances of $J$ range from about 20 to 800. But the variances of AoVM in the rest of the fields 
range from about 20 to 400. The largest clusters in the rest 34 fields also have a higher central value of AoVM than other fields generally. 
In the covariance matrix of the largest cluster, the covariance between $\eta$ and $J$ has the smallest negative value from about $-6$ to 0. 
Only minor fraction of fields show the smallest negative covariance between $\eta$ and AoVM. As shown in Paper I, a substantial number  
of light curves exhibit negative correlations between $\eta$ and $J$, or between $\eta$ and AoVM in the NSVS data, 
implying systematic properties of the NSVS data.

\section{Separation of variable candidates}\label{sec:sep}

One disadvantage of clustering algorithms is that there is no useful validation process for 
clustering results as in other unsupervised learning methods. Our approach using the infinite GMM also shares 
this problem with other clustering methods. Therefore, there is no one absolutely right way to define 
a boundary between variable and non-variable objects in multi-dimensional space with the clustering results. 
In many situations, 
the selection rule of variable objects can be limited by practical issues such as the number of objects 
which can be investigated further in follow-up studies. Here, we suggest three possible ways to separate 
variable objects from non-variable objects by using clustering results.

As suggested in Paper I, first, we can select as variable candidate objects which are not included in 
large groups \footnote{We warn that this approach does not work if the most dominant fraction of light 
curves corresponds to variable sources.}. 
Figure \ref{fig:gmm} shows that the largest group in each field includes a different 
number of objects. Therefore, defining objects in the largest group as non-variable objects produce 
different numbers of variable candidates in each field. Moreover, if there were any systematic patterns 
in light curves, or if the majority of non-variable light curves is not well described by a 
multivariate Gaussian distribution, data in each field would make the second or third large groups 
include non-variable objects or objects affected by the systematic patterns in their light curves. 
Basically, the existence of small clusters, i.e., the exact number of clusters, is not an 
important factor to select variable candidates.

\begin{figure}
\plotone{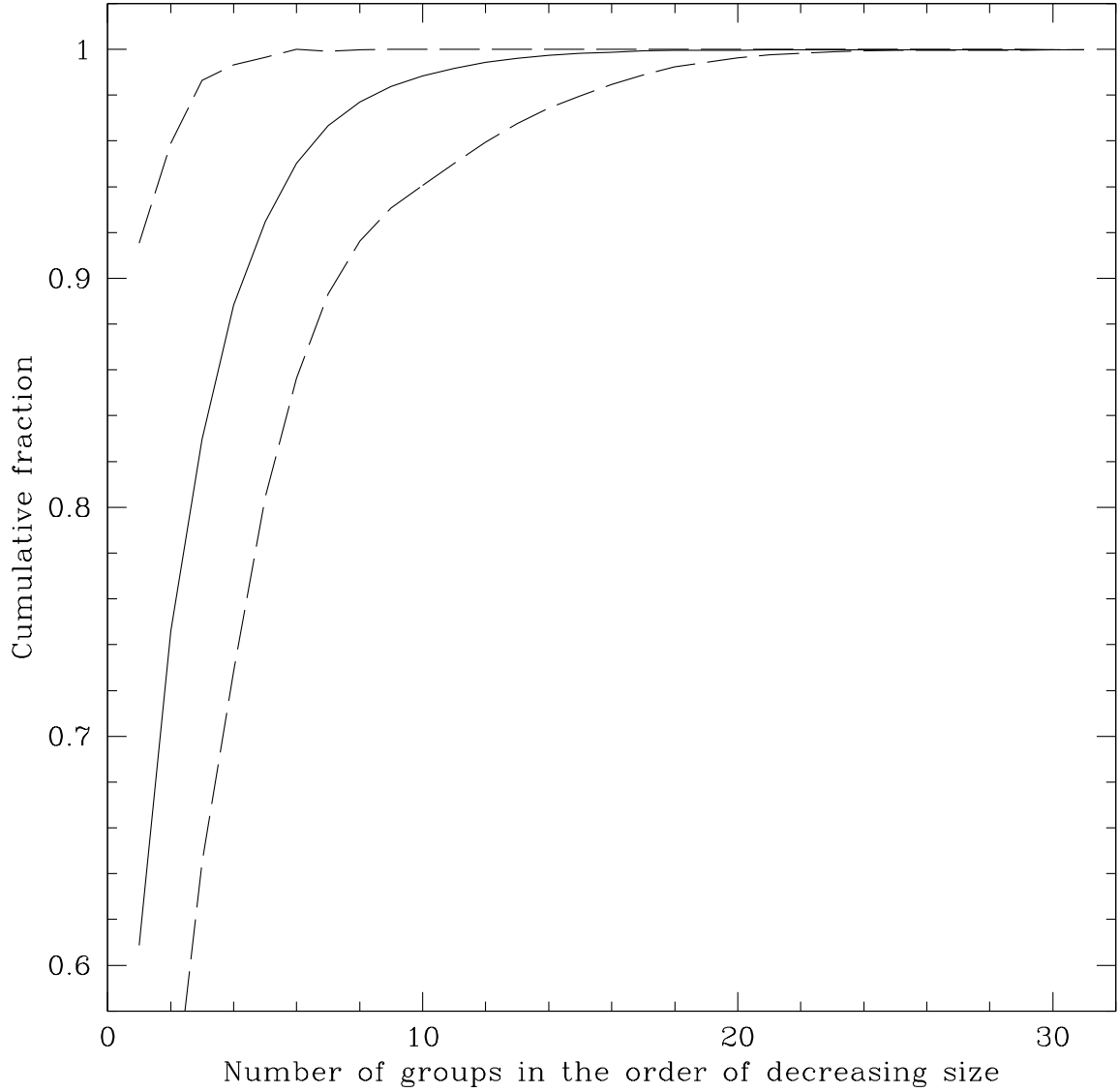}
\caption{Cumulative fraction of objects included in groups. The median fraction as a function of the number of groups is presented with 
the solid line, while dashed lines describe the minimum and maximum fractions. For example, top four large groups 
explain about 89\% of light curves as the median fraction. The median fraction of about 99\% is found with 
top ten large groups.}
\label{fig:large_groups}
\end{figure}

Figure \ref{fig:large_groups} presents the change of cumulative fractions of data included in groups 
as the number of large groups increases. Since we find only six clusters in the field 147d, 
the fraction of data included in the top four groups is about 99\%. However, the median fraction of data included in 
the top four large groups is about 89\%. To select variable candidates from 
minor groups in each field, one could adopt a 90\% cut of the cumulative fraction and avoid light curves 
included in large groups.

\begin{figure}
\plotone{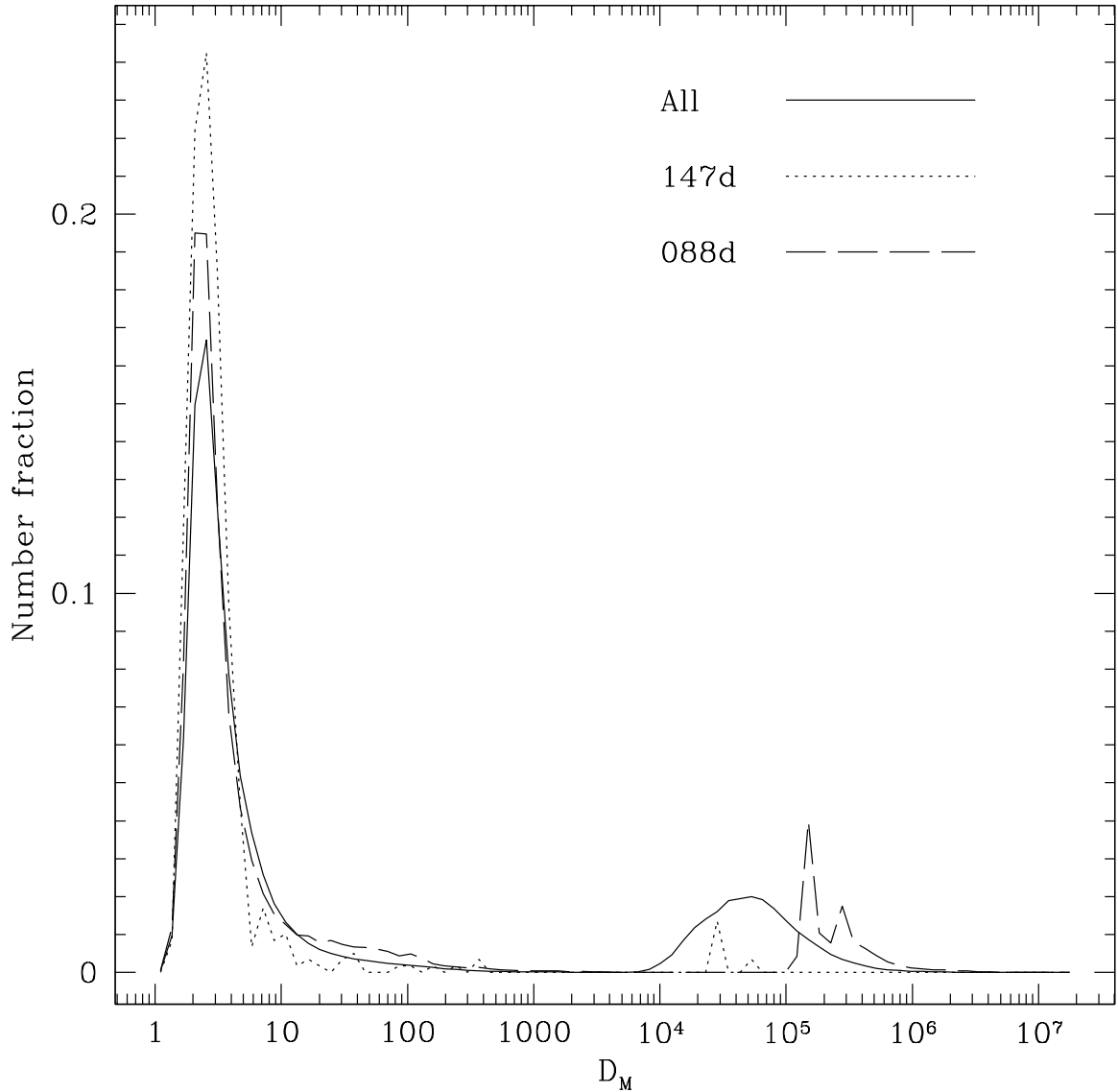}
\caption{Distributions of $D_{M}$. The peak of the distributions around $D_{M} \sim 2$ corresponds to a mode value 
of $D_{M}$ for a multivariate normal distribution \citep{ververidis08}. Compared with the distribution for all objects 
(solid line), the field 147d (dotted line), which has the smallest number of light curves, 
and the field 088d (dashed line), which has the largest number of light curves, in our samples 
are more dominated by objects around the largest cluster in the eight-dimensional space.
}
\label{fig:D_M}
\end{figure}

Another possible approach is using distances of objects from the center of the largest group in the eight-dimensional 
space spanned by the variability indices. In Paper I, we introduced the Mahalanobis distance from the 
largest cluster:
\begin{equation}
D_{M} = \sqrt{(\bf{x} - \bf{\mu}_{0})^{T}\Sigma_{0}^{-1}(\bf{x} - \bf{\mu}_{0})},
\label{eq:D_M}
\end{equation}
where the center $\bf{\mu}_{0}$ and covariance matrix $\Sigma_{0}$ of the largest cluster are used 
with the position of an individual object $\bf{x}$ in the eight-dimensional space. Unlike a commonly used 
Euclidean distance, $D_{M}$ depends on $\Sigma_{0}$ which describes how broadly the objects in the largest 
cluster disperse \citep{sharma09}. Objects with large distances can be considered as variable objects because they 
have strong dissimilarity from the dominant fraction of light curves \citep[e.g.,][]{jolion91}. In Figure 
\ref{fig:D_M}, the distribution 
of $D_{M}$ for all objects with respect to the largest cluster in each field shows a peak of 
the distribution around $D_{M} \sim 2$ corresponding to 
the mode value of the beta distribution which is expected for the distribution of $D_{M}$ \citep{ververidis08}. 
It also shows another concentration of objects between 
$D_{M} \sim 10^{4}$ and $10^{6}$, implying the possible signature of a large number of variable candidates 
or any systematic patterns hidden in the NSVS light curves. 
This feature varies strongly in each field as shown for the fields 088d and 147d in Figure 
\ref{fig:D_M}.

We can derive the cut of $D_{M}$ that includes a specific fraction of members in the 
largest cluster by integrating the multivariate Gaussian distribution with the largest cluster's 
center $\bf{\mu}_{0}$ and covariance matrix $\Sigma_{0}$. For example, the $b$\% cut of $D_{M}$ can be found with 
\begin{equation}
\int_{\bf{x}:D_{M}(\bf{x})<D_{M}^{\rm cut}} p(\bf{x} \vert \bf{\mu_{0}}, \Sigma_{0}) d\bf{x} = b,
\label{eq:D_M_cut}
\end{equation}
where $p(\bf{x})$ is a multivariate Gaussian distribution. 
Practically, the integration can be estimated by using the Monte Carlo method 
\citep{chen06}. Each field has a different value of 
$D_{M}^{\rm cut}$ for the same probability cut. For example, the 99\% cut in the field 088d is $D_{M}^{\rm cut} ~=~ 5.634$, 
while $D_{M}^{\rm cut} ~=~ 5.105$ corresponds to the 99\% cut in the field 147d. Since the largest cluster has the sharper concentration of 
$D_{M}$ in the field 147d than in the field 088d (see Figure \ref{fig:D_M}), $D_{M}^{\rm cut}$ in the field 147d is 
smaller than that in the field 088d.

These two approaches have their own different problems. Avoiding large groups does not guarantee that minor groups are 
well separated from the largest group. In a multi-dimensional space, a few minor groups can be close to the 
large groups with a small $D_{M}$. When selecting objects with $D_{M} > D_{M}^{\rm cut}$ as variable candidates, those objects 
with a large $D_{M}$ can form large groups representing systematic observational features in each field. 
Although Figure \ref{fig:D_M} implies that members of most large groups have a small $D_{M}$, using only one 
method can cause more contamination of non-variable objects or objects dominated by systematic patterns in each observation 
field.

\begin{figure}
\plotone{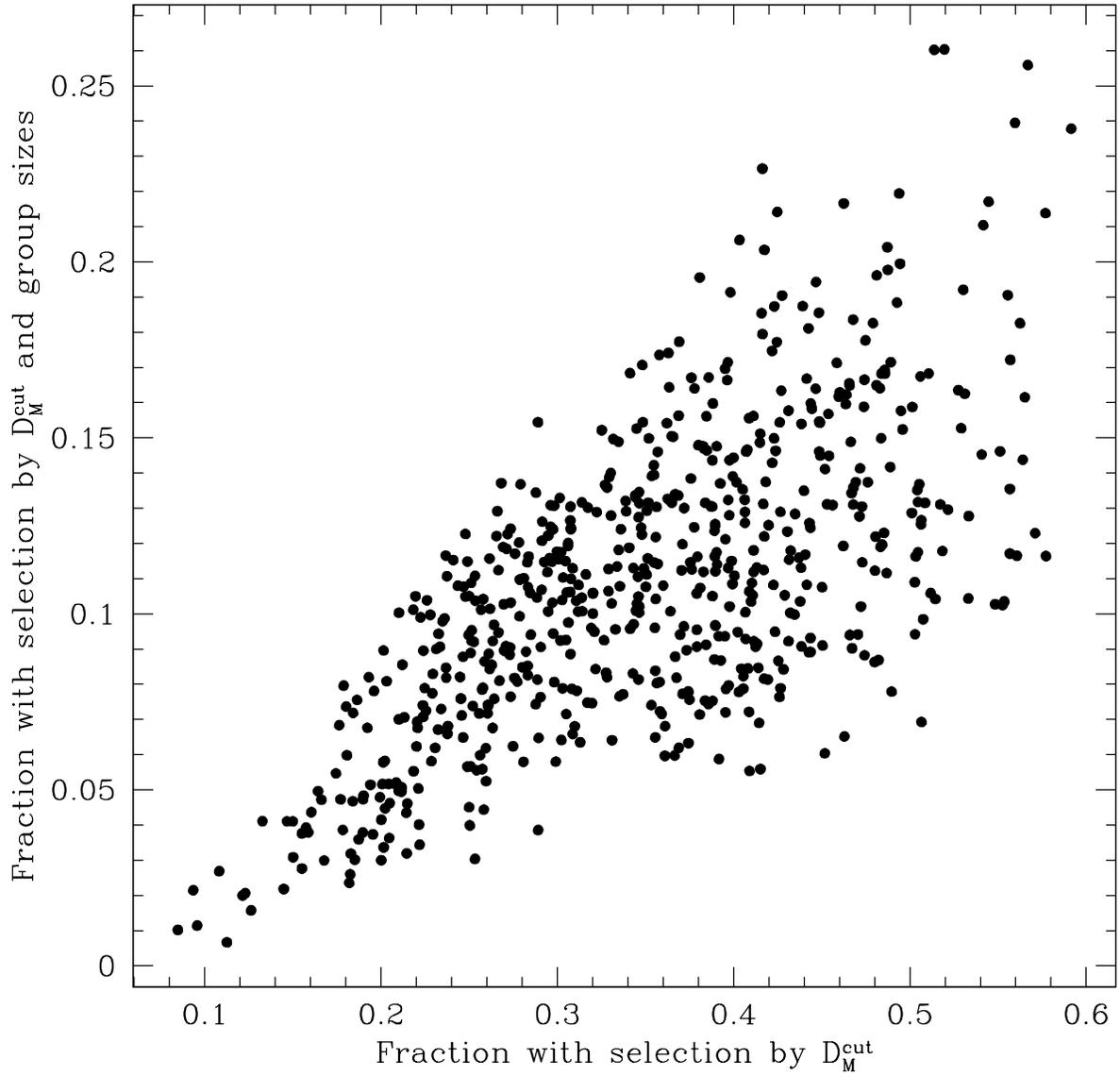}
\caption{Fraction of objects selected as variable candidates. 
As found in the different ranges of the horizontal axis and the vertical axis, 
the size of variable candidates can be decreased significantly by 
combining the constraints of $D_{M}^{\rm cut}$ and group sizes together.
}
\label{fig:both}
\end{figure}

We can define variable candidates conservatively by combining both methods. For example, we find 
objects having the larger $D_{M}$ than the $D_{M}^{\rm cut}$ of the 99\% cut and not being included in the top four large groups. 
Figure \ref{fig:both} represents the number fraction of objects selected by this conservative method compared to results of 
selecting objects with the $D_{M}^{\rm cut}$ of the 99\% cut. We use this conservative selection of variable candidates in Section 
\ref{sec:var}. 
Because the top four large groups are closely located in the eight-dimensional 
space generally, excluding the members of the top four large groups from the variable candidates usually 
determines the size of variable candidates.

\section{Database}\label{sec:web}

We provide the results of our variability analysis and clustering for all sample light curves online. 
The database stores 
the eight variability indices, the cluster identification number in each field, $D_{M}$ from the largest cluster for every light curve as well as 
the basic information of the light curves such as the NSVS object id and coordinates. 
The database is supplemented with 
the number of light curves analyzed, the number of groups found by clustering, 
the number fractions of group members, and the $D_{M}$ cuts of 99\%, 95\%, and 90\% cuts from the largest group in each field. 
Therefore, users of the database can select variable candidates by using these clustering results with their own selection 
rules of variable candidates.

The database is also supplemented by association to other astronomical catalogs. All objects analyzed 
are cross-matched to the SIMBAD database, 2MASS All-Sky Catalogs of Point Sources \citep{2mass}, 
the photometric catalog of the Sloan Digital Sky Survey (SDSS) Data Release 7 \citep{sdss}, and 
the photometric catalog of the {\it GALEX} GR4/GR5 Data Release \citep{galex} with 
the NSVS coordinates and a search radius of 6$''$. When multiple objects are matched within the search radius, 
the nearest match is associated with the NSVS object. In addition to these catalogs, we also present 
matching results of the NSVS coordinates to {\it IRAS} \citep{helou88,moshir90} and AKARI \citep{akari} catalogs as 
separate files online. 
This association to other catalogs can be used to 
identify the morphology of the NSVS objects as galaxies or stars, to find astronomical types of known objects, 
to check blending effects with neighbor objects, and to estimate colors of objects as we show in the following section. 
In particular, because the spatial 
resolution of the NSVS data is much worse than the SDSS and 2MASS catalogs, the morphological 
information from these catalogs can help database users to avoid blended objects and less precise 
photometry in the NSVS catalog.

The database can be accessed through the Web interface\footnote{\url{http://stardb.yonsei.ac.kr}}. Searching the light 
curves and their variability analysis is possible with equatorial coordinates and a search radius given by users. 
In particular, the simple cone search interface\footnote{\url{http://stardb.yonsei.ac.kr/conesearch/nsvs_conesearch.php}}
is also provided for compatibility with the Virtual Observatory environment \citep{williams08}. We plan to provide 
the basic components of the database in Vizier\footnote{\url{http://vizier.u-strasbg.fr/viz-bin/VizieR}} too.

\section{Properties of variable candidates}\label{sec:var}

In this section, we examine properties of variable candidates which are not 
included in the top four groups and have $D_{M}$ larger than the 99\% cut. With these conservative 
selection criteria, we find 1,840,310 light curves as possible variable candidates. 
We emphasize that this selection of variable candidates is highly conservative. When we select variable 
candidates simply with $D_{M}$ larger than the 95\% or 99\% cut, 
the total number of variable candidates is 6,640,387 or 5,826,587, which is about 3.6 or 
3.1 times more than the number of variable candidates selected with the conservative definition, respectively. 
Meanwhile, when we 
select objects not included in the top four groups as variable candidates, the 
number of candidates is 1,918,580, which is similar to the result of the conservative selection. 
But we remind that this selection of variable candidates is still affected by the intrinsic limits 
of the NSVS data such as blending effects and completely different uncertainty properties of 
photometry for extended objects compared to stellar objects.

\subsection{Known objects and new candidates}\label{sec:simbad}

We search all NSVS samples with the SIMBAD database to recognize any known objects and their basic properties such as
well-known names. As explained in Section 4, the nearest SIMBAD object around the coordinates of the NSVS objects is 
retrieved with a search radius of 6$''$. Since the NSVS data do not have information about morphological classification 
such as galaxy and star, the auxiliary information of the SIMBAD database is useful to sort out spurious variability of extended 
objects. About 6\% of the NSVS objects are matched with at least one SIMBAD object. But this fraction is 
affected by the precision of the NSVS coordinates.

Based on the object classification given in the SIMBAD database\footnote{\url{http://simbad.u-strasbg.fr/simbad/sim-display?data=otypes}}, 
we find that 16,061 NSVS light curves correspond to known or suspected variable stars. Considering only 
variable candidates with our conservative selection criteria, the number of known or suspected variable stars is 
11,080 among our variable candidates. Finally, excluding known or suspected variable stars as well as 
known galaxies in the SIMBAD database, 
the number of new variable candidates is 1,824,123, with our conservative selection. Hereafter, 
our investigation of variable candidates is limited to these 1,824,123 NSVS objects. We note that uncertain coordinates 
in either NSVS or the SIMBAD databases can cause us to miss some known galaxies and known or suspected variables. The classifications in the SIMBAD databases might not be as complete as other catalogs of 
variable stars such as the AAVSO International Variable Star 
Index (VSX)\footnote{\url{http://www.aavso.org/vsx/}} \citep{watson06} as we discuss in 
Appendix.

Figure \ref{fig:missing_exam} shows light curves of six example NSVS objects which are not included in 
our conservative selection of variable candidates, but which are known or suspected variables in the 
SIMBAD database. These examples are among 11 objects, which are not included in our candidates, in the NSVS 
field 112a corresponding to a part of the constellation Aquila. If we selected variable candidates 
as objects with $D_{M}$ larger than the 95\% cut, four objects, including \object{GSC 00490-04680} 
\citep{bernhard00} and NSV 12564 \citep{kinnunen00} in Figure \ref{fig:missing_exam}, 
would be included in variable candidates among the 11 objects. The field 064d misses the largest number 
of known or suspected variable stars (236 objects) with our conservative selection of candidates. However, 
its fraction is only 0.6\% with respect to the total number of objects that are analyzed by our clustering method. 
We note that most these missing objects are identified as variable objects in the {\it Kepler} field by HATNET 
which uses image subtraction method \citep{hartman04}.

\begin{figure}
\plotone{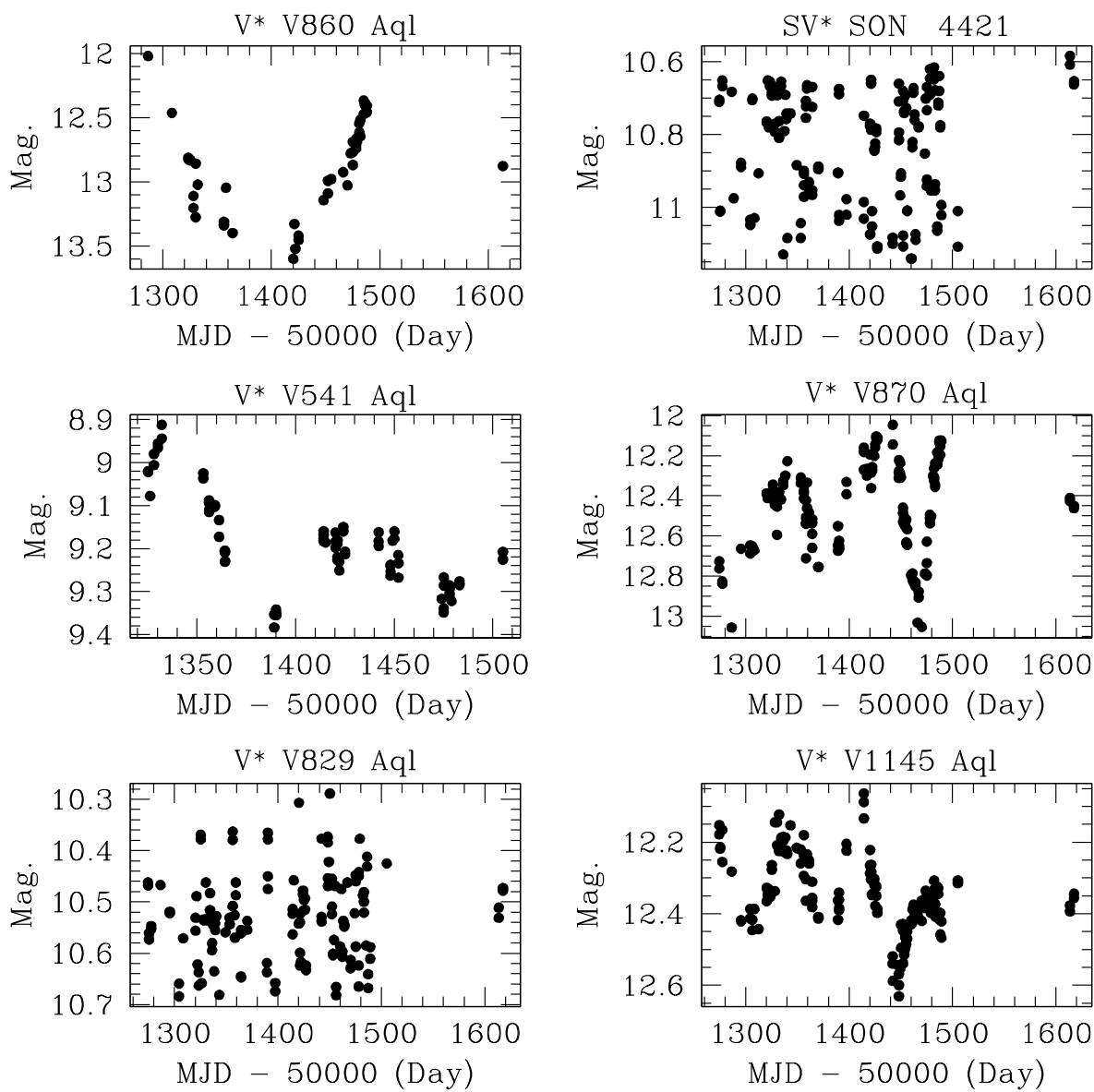}
\caption{Example light curves for known or suspected variable objects included in our conservative variable candidates. 
The SIMBAD names of the objects are given in the top of each panel.}
\label{fig:found_exam}
\end{figure}

Figure \ref{fig:found_exam} presents example light curves of known or suspected variable stars matched with the 
SIMBAD database and included in our conservative selection of variable candidates for the NSVS field 112a. We 
find 88 known or suspected variable stars among our 6,389 variable candidates in the field 112a, 
corresponding to about 1\%.

\begin{figure}
\plottwo{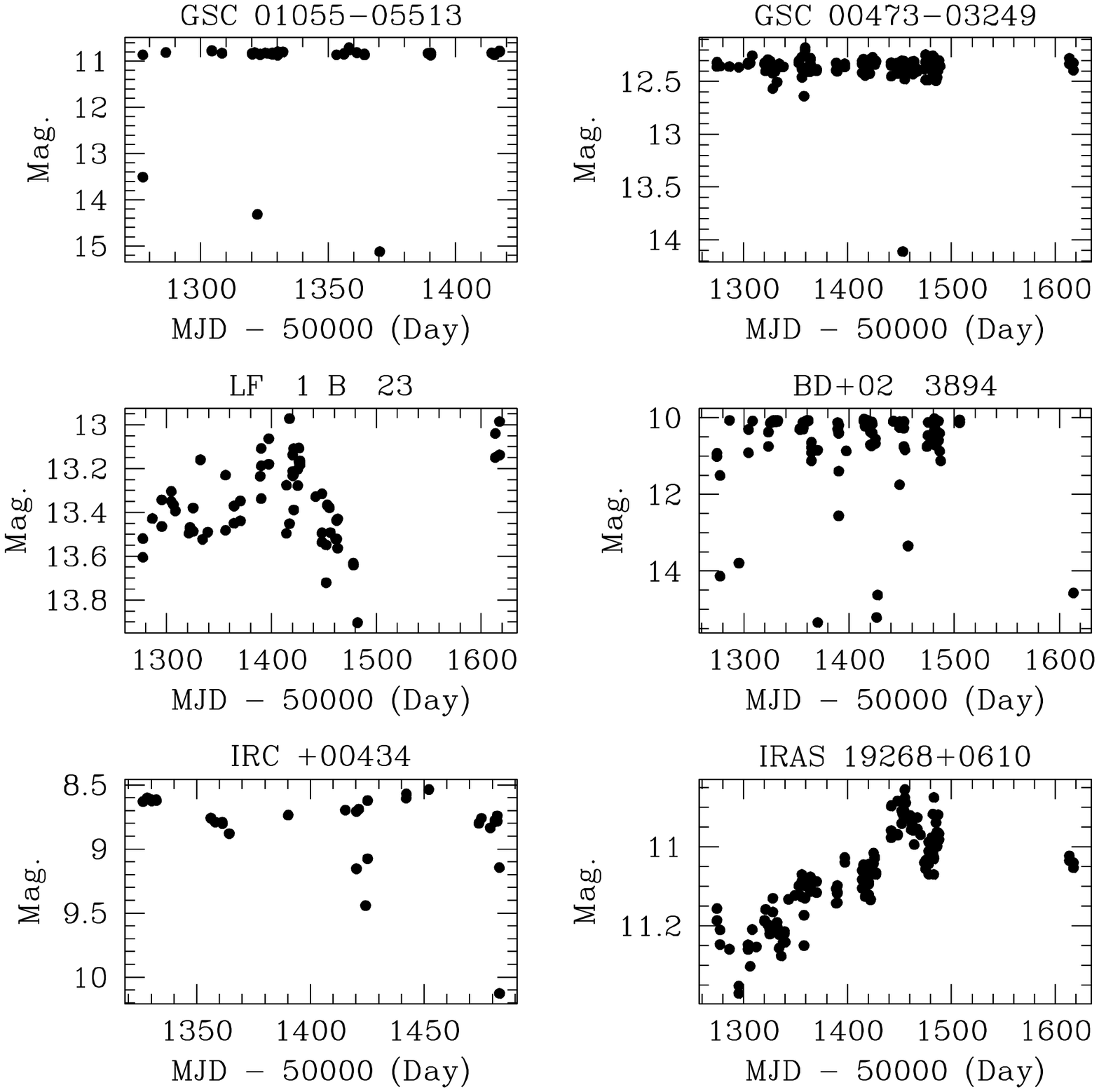}{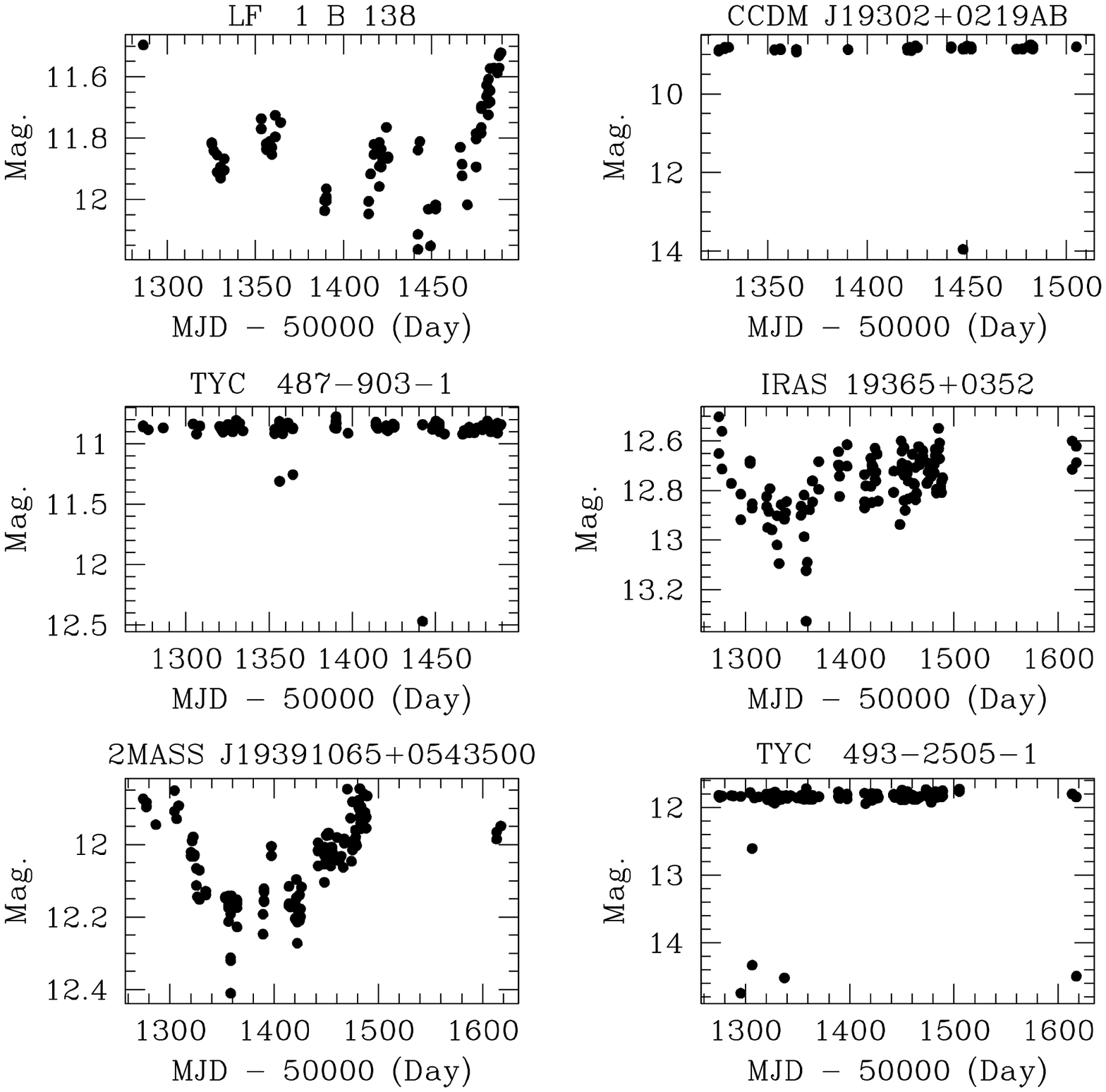}
\caption{Example light curves of variable candidates matched to SIMBAD objects of non-variable stars. 
Each light curve corresponds to the SIMBAD object with the name given in the top of each panel. Although 
IRC +00434 and 2MASS J19391065+0543500 are not found as known variable sources in the SIMBAD database, 
these objects are found in the VSX catalog \citep{pojmanski02,usatov08b} as we explain in Appendix.
}
\label{fig:new_exam}
\end{figure}

In Figure \ref{fig:new_exam}, we show 12 examples of new variable candidates which are matched to any kind of known 
objects except variable stars in the SIMBAD database. \object{CCDM J19302+0219AB} is not typed as variable stars in the SIMBAD 
database. But this object is a known system of double stars \citep{dommanget94} which might have been affected 
by blending in the NSVS data. 
\object{2MASS J19391065+0543500} is also not included as a variable star in the SIMBAD database. But \citet{usatov08b} 
suggest that this object is a red asymptotic giant branch (AGB) variable star, supporting that the light curve shown in Figure \ref{fig:new_exam} 
exhibits true light variation. These examples show that our conservative selection of variable candidates 
can catch real variable objects.

We also compare our variable candidates to those found by others using the NSVS data. For example, 
785 RR Lyrae candidates were reported already \citep{wils06}. These 781 objects are included in our conservatively selected 
candidates. But all 785 objects are found when we select variable candidates as objects having  larger than 
the 99\% $D_{M}$ cut. When we select variable candidates with the 95\% $D_{M}$ cut, all new $\beta$ Lyrae and Algol-type variable 
candidates from \citet{hoffman08} are recovered with our method if they are included in our original NSVS samples. 
But we recover 95\% of them with the conservative selection of variable candidates. About 
4700 variable candidates of other kinds such as $\delta$ Scuti stars and Cepheid objects were also studied with the NSVS data 
\citep{hoffman09}. Again, our selection with the 95\% $D_{M}$ cut recovers most variable candidates except for 
objects which have different light curves due to different definitions of good photometric data and different systematic 
patterns in light curves \citep{hoffman09}. 
These other studies use simple rules such as 0.1 mag dispersion of light curves, which are conservative selection methods 
for specific types of variable candidates. Therefore, the number of variable candidates is much larger in our approach than in others. 

\subsection{{\it IRAS} sources}

For the conservative selection of variable candidates, we find {\it IRAS} sources which are spatially matched to the NSVS coordinates 
with a search radius of 6$''$ in the {\it IRAS} point source catalog \citep{helou88} and the {\it IRAS} faint source 
catalog \citep{moshir90}. Among all NSVS samples, 31,852 light curves have the matching {\it IRAS} sources. Considering 
the {\it IRAS} sources only for our conservative variable candidates, the number is 12,987, which is about 41\%.

We derive two colors of the {\it IRAS} sources by using the {\it IRAS} photometric flux 
at 12, 25, and 60 ${\rm \mu m}$ ($F_{\rm 12 \mu m}$, $F_{\rm 25 \mu m}$, and $F_{\rm 60 \mu m}$). 
The conventional definition of the {\it IRAS} colors \citep[e.g.,][]{vanderveen88,oliver01,sevenster02} is 
\begin{equation}
C_{12/25} = 2.5 {\rm log} (\frac{F_{\rm 25 \mu m}}{F_{\rm 12 \mu m}}); 
C_{25/60} = 2.5 {\rm log} (\frac{F_{\rm 60 \mu m}}{F_{\rm 25 \mu m}}) 
\label{eq:IRAS_color}
\end{equation}
where we do not apply any color corrections to the fluxes.

\begin{figure}
\plotone{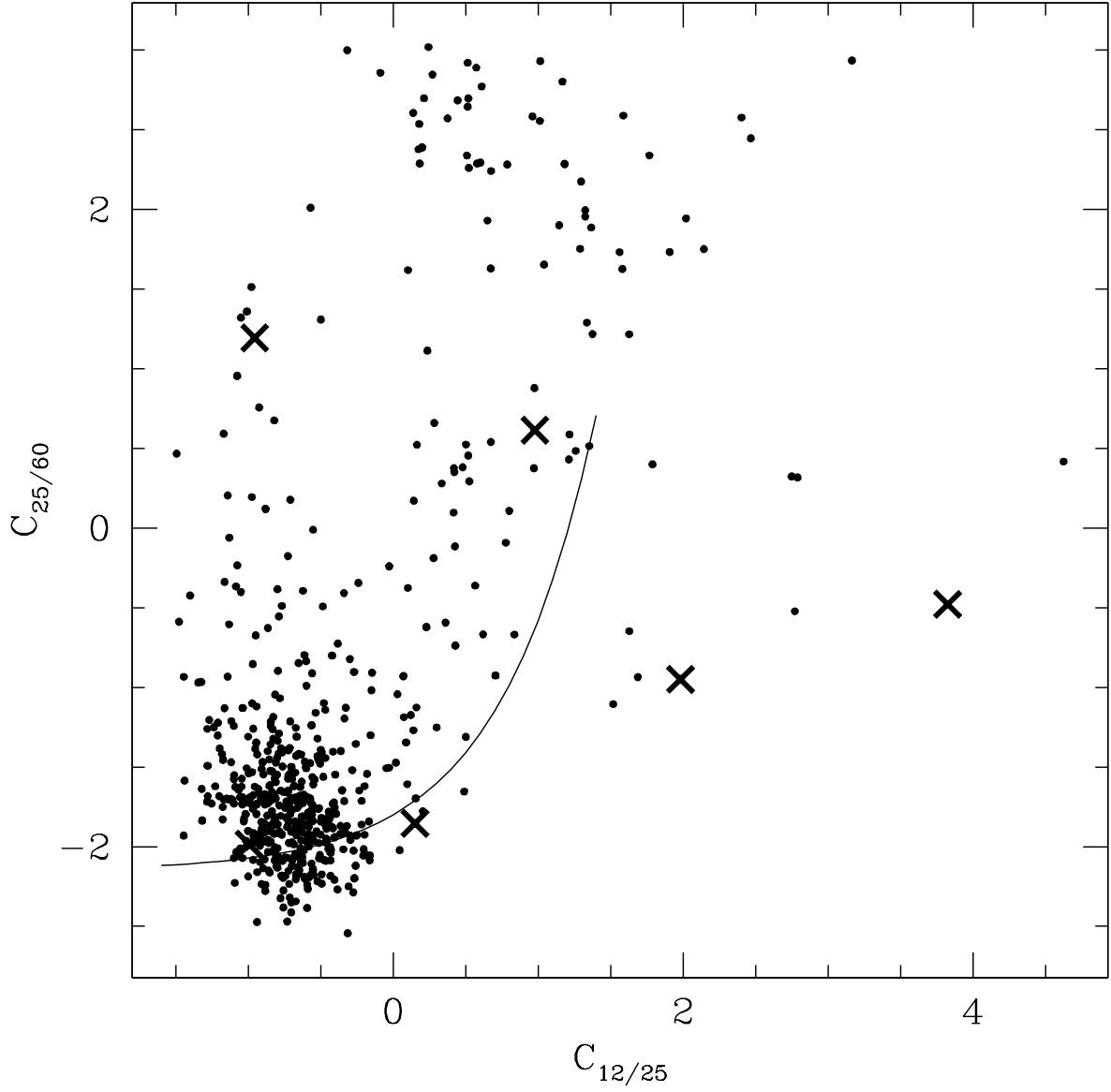}
\caption{{\it IRAS} two-color diagram of variable candidates. 
The solid line represents the color of oxygen-rich Mira variables and variable OH/IR stars 
from \citet{vanderveen88}. Cross symbols correspond to objects shown in Figure \ref{fig:iras_lc}.
}
\label{fig:iras_cc}
\end{figure}

The {\it IRAS} colors have been commonly used for classification of infrared sources. In particular, the two-color diagram 
like Figure \ref{fig:iras_cc} helps us understand what kind of variable candidates show variability which is relevant to the  
late stage of stellar evolution such as AGB stars with evolved circumstellar dust \citep{zuckerman87,vanderveen88,kwok97,
ramos-larios09}. In Figure \ref{fig:iras_cc}, we plot the colors of {\it IRAS} sources 
with the quality number $Q = 3$ at all three wavelengths 12, 25, and 60 ${\rm \mu m}$ 
\citep{helou88}. AGB stars dominate colors of $C_{25/60} < -0.3$, while planetary nebulae and 
young stellar objects dominate $-0.3 < C_{25/60} < 0.4$ and $0.4 < C_{25/60}$, respectively \citep{jackson02}.

\begin{figure}
\plotone{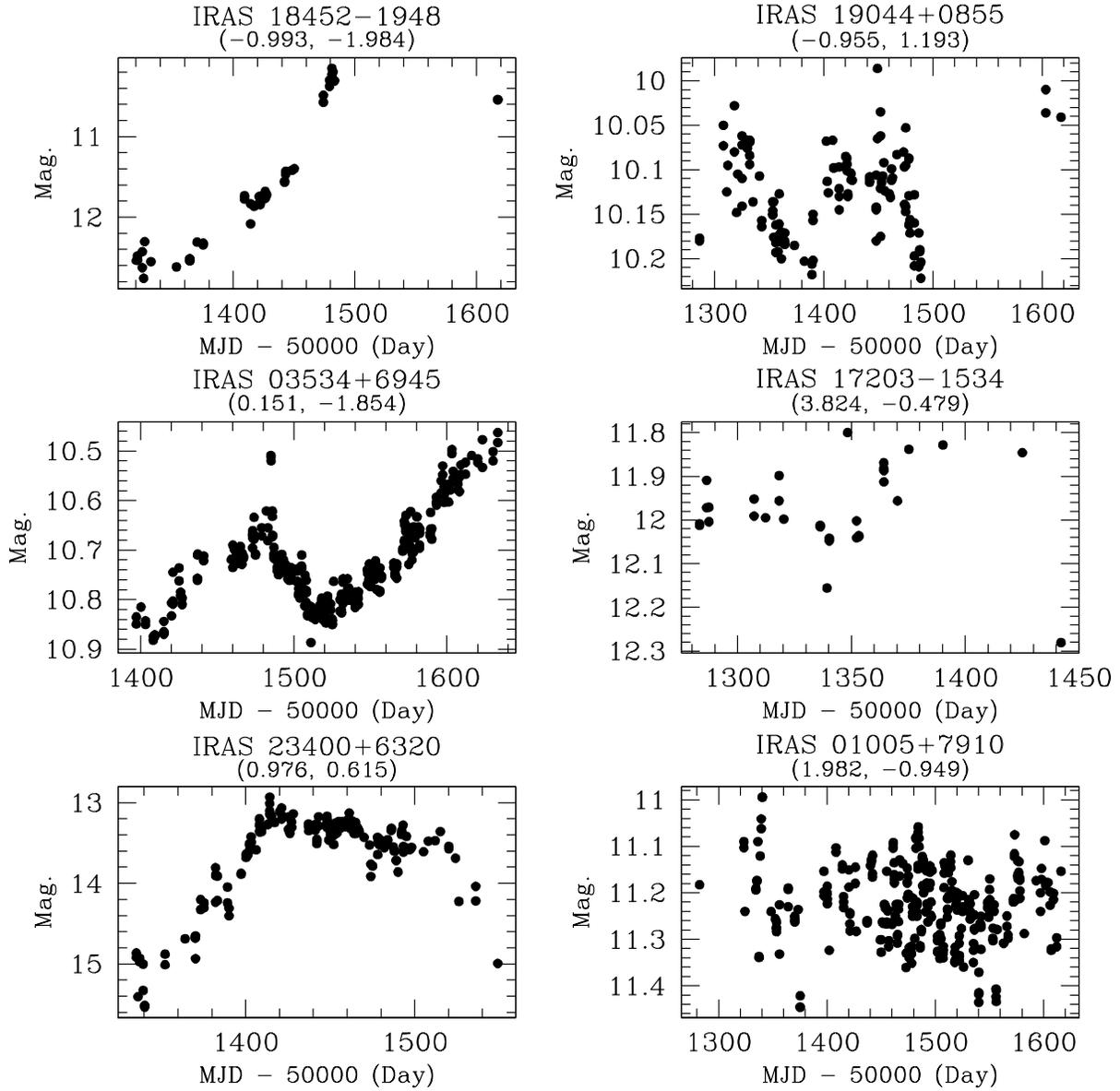}
\caption{Example light curves of variable candidates that are {\it IRAS} sources. The 
{\it IRAS} designations and ($C_{12/25}$, $C_{25/60}$) are presented in the top of each panel. 
Objects on the left column have colors near the curve presented in Figure \ref{fig:iras_cc}.
}
\label{fig:iras_lc}
\end{figure}

In Figure \ref{fig:iras_lc}, we show example light curves of variable candidates that have the corresponding 
{\it IRAS} identifications. Although these {\it IRAS} sources are not known variable stars in the SIMBAD database, 
some of them have been investigated 
in various ways without variability information. \object{IRAS 03534+6945} (NSVS 513536) and \object{IRAS 23400+6320} (NSVS 1487299) 
were found as H-$\alpha$ emitting stars in \citet{stephenson86} and \citet{coyne83}, respectively. These sources 
are also found as a possible variable sources in the VSX catalog (see Appendix). 
\object{IRAS 17203-1534} (NSVS 16483061) and \object{IRAS 01005+7910} (NSVS 262162) are post-AGB stars which are sub-classified as 
hot post-AGB stars and high galactic latitude supergiants \citep{szczerba07}, respectively. \object{IRAS 01005+7910} has 
also been observed with the {\it Hubble Space Telescope} which found nebulae around it \citep{siodmiak08}.

\subsection{AKARI sources}

Bright objects included in this paper are expected to be included in AKARI observations which have been conducted with 
two instruments Far-Infrared Surveyor (FIS) and Infrared Camera \citep[IRC;][]{akari}. Both instruments produced all-sky 
source catalogs which are much deeper and spatially better resolved than {\it IRAS} \citep{akari_cat}. 

We match our NSVS samples to AKARI/IRC All-Sky Point Source Catalog  \citep[Version 1.0;][]{akari_irc,akari_irc_cat} and 
AKARI/FIS All-Sky Survey Bright Source Catalog \citep[Version 1.0;][]{akari_fis_cat}, with a search radius of 6$''$. 
The numbers of our NSVS samples matching to the AKARI objects are 267,732 and 8,742 for the IRC and FIS catalogs, respectively. 
This matching rate with the IRC catalog is much higher than that with the {\it IRAS}.

\begin{figure}
\plotone{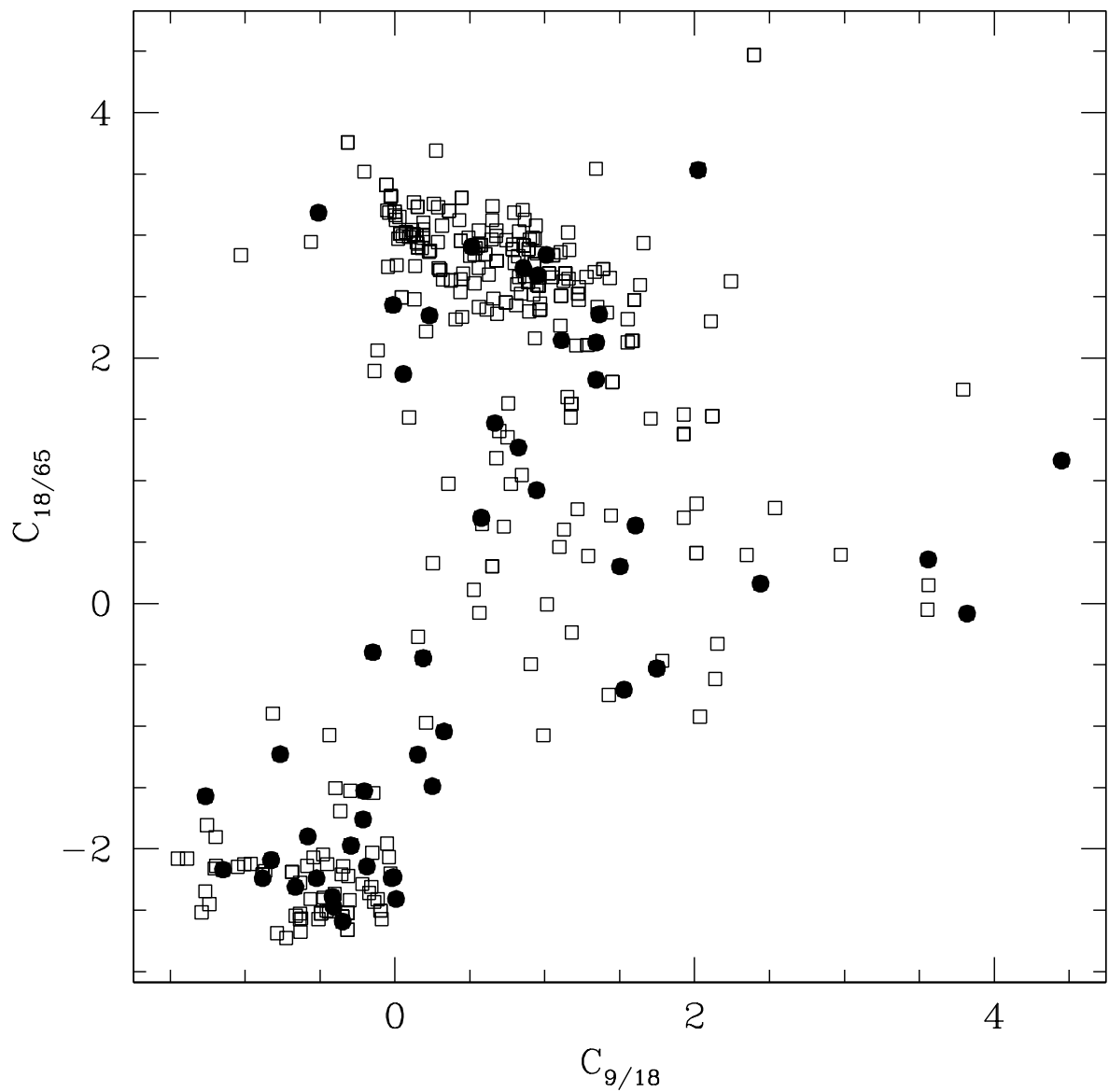}
\caption{AKARI two-color diagram of variable candidates and other objects. Following the definition of the 
{\it IRAS} colors given in Equation (\ref{eq:IRAS_color}), AKRAI colors are defined by using fluxes at its IRC 9, 18 $\mu$m 
and FIS 65 $\mu$m. Filled circles represent new variable candidates selected with our conservative selection, while empty 
squares correspond to non-variable objects or objects that are known variable stars or galaxies in the SIMBAD database.
}
\label{fig:akari_cc}
\end{figure}

Figure \ref{fig:akari_cc} shows a color -- color diagram of AKARI fluxes like Figure \ref{fig:iras_cc}. In the plot, we show 
objects as identified as point sources with only good photometric observations of AKARI IRC at 9 and 18 $\mu$m, 
and FIS at 65 $\mu$m, which are found with IRC photometric flags of conditions q\_S09 $=$ 3, f09 $=$ 0, X09 $=$ 0, q\_S18 $=$ 3, 
f18 = 0, X18 = 0, and with FIS photometric flags of conditions q\_S65 $>$ 1. The number of the NSVS objects with the good 
photometric data is 374, while only 54 AKARI point-source objects with good photometric data correspond to the variable candidates 
selected conservatively. The distribution of AKARI colors is similar to that of {\it IRAS} colors because of the similar wavelength 
ranges of the observation bands.

\begin{figure}
\plotone{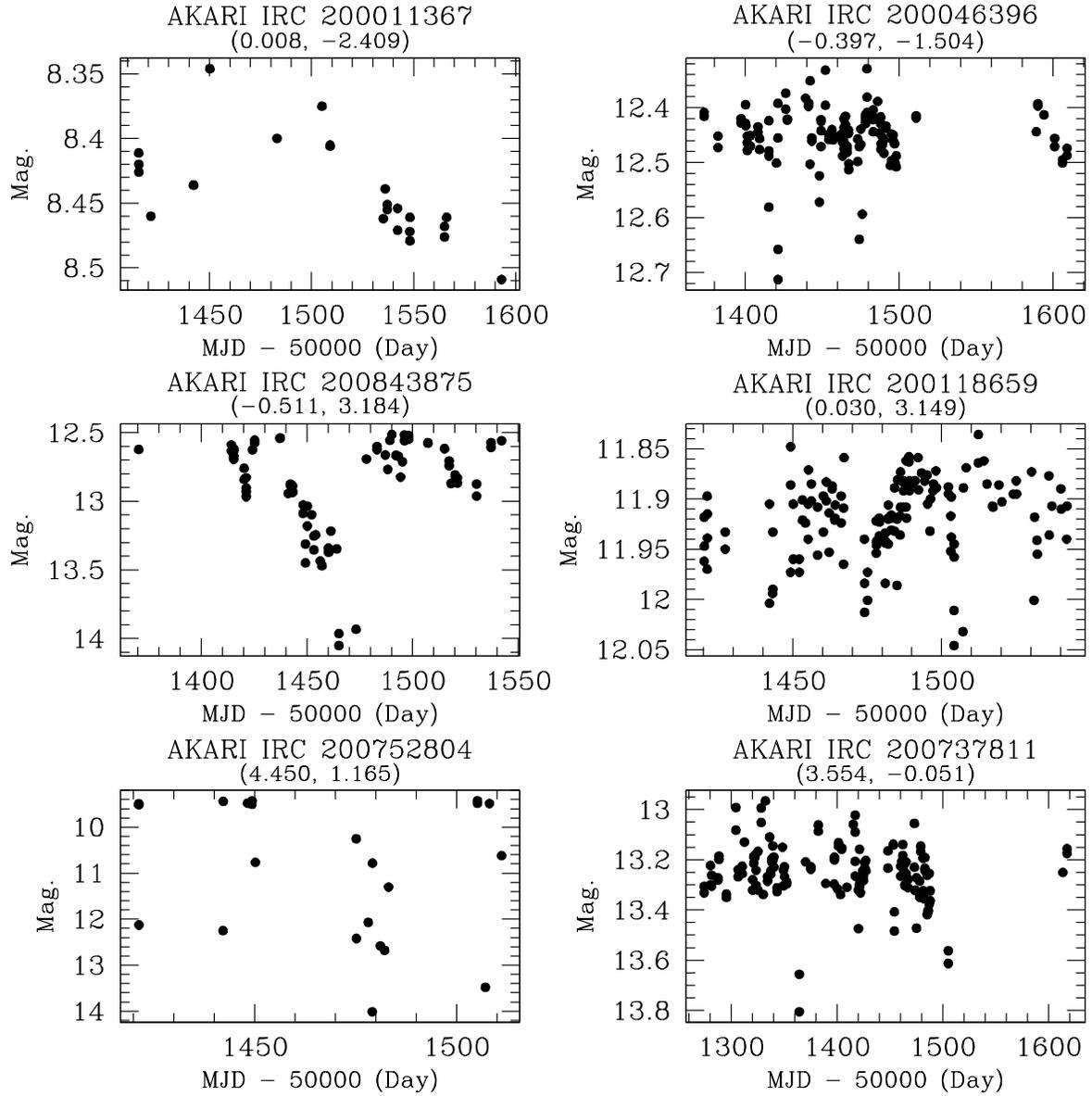}
\caption{Example light curves of variable candidates (left) and non-variable objects (right) that are AKARI sources. 
The AKARI designations in the IRC catalog and (${\rm C_{9/18}}$, ${\rm C_{18/65}}$) are presented in the top of each panel.
}
\label{fig:akari_lc}
\end{figure}

As we find with the {\it IRAS} colors, the variable candidates with the corresponding AKARI objects might be long-period late-type stars 
\citep{ita10}. 
Figure \ref{fig:akari_lc} shows example light curves of the variable and non-variable candidates in the 
SIMBAD database. AKARI IRC 200011367 (NSVS 1713088) corresponds \object{TYC 3668-112-1}, which is also an {\it IRAS} object. Even though 
this light curve does not have many observed data points, the light curve is selected as a possible variable candidate in the NSVS field 
013b. AKARI IRC 200843875 (NSVS 3393876) is close to \object{IRAS 22164+6427}, which might be the same object. 
The light curve of AKARI IRC 200752804 does not have many observed data points. But its variation seems reasonable because of the 
fact that it is a post-AGB star or a protoplanetary nebula, corresponding to \object{HD 331319} and \object{IRAS 19475+3119}. 
The light curves of the other three objects presented in Figure \ref{fig:akari_lc} do not correspond to any known objects in the 
SIMBAD database, and are not selected as variable objects.

\subsection{2MASS and SDSS photometry}\label{sec:2mass_sdss}

Near-infrared (NIR) colors are also commonly used to identify basic properties of stars and to separate non-stellar 
objects such as quasars. We match all variable candidates to the 2MASS All-Sky Catalog of Point Sources \citep{2mass} 
with a search radius 6$''$. A total of 1,439,381 variable candidates have corresponding 2MASS sources. 
Hereafter, 2MASS photometry data are given in Vega system.

\begin{figure}
\plotone{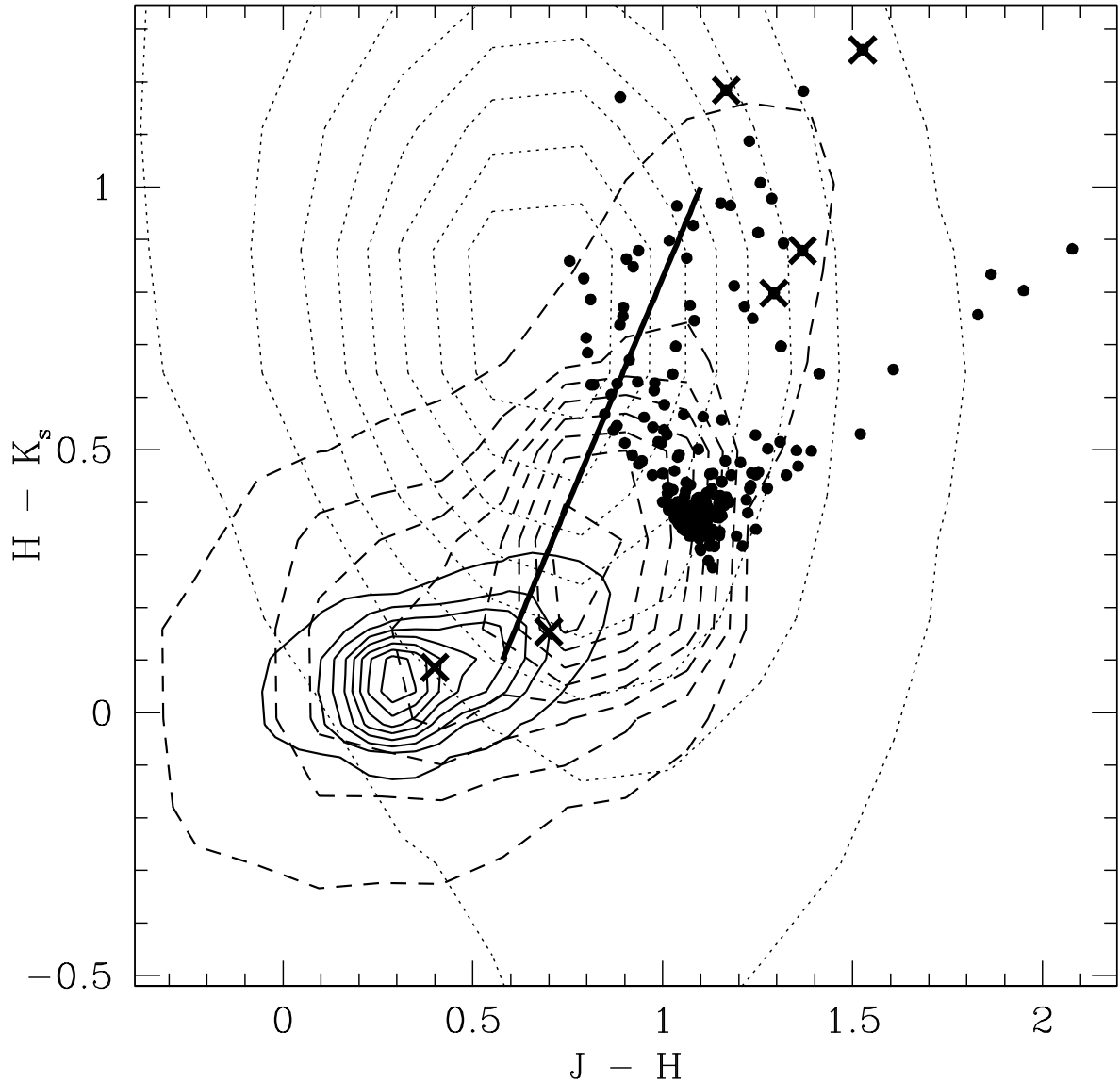}
\caption{2MASS color-color diagram of variable candidates. The color distribution of variable 
candidates (solid-line contours) is similar to the overall color distribution of 
ordinary stars \citep{covey07}, while observed colors of 6658 QSOs with redshifts $>$ 0.3 from \citet[][dotted-line contours]{veron-cetty06} 
are not similar to those of variable candidates. The color distribution of 
known pulsating variable stars in LMC and SMC \citep[dashed-line contours;][]{ita04} implies that variable 
candidates redder than ($J - K_{s}$) = 1.4 (dots) might be variable carbon-rich stars. The loci of classical 
T Tauri stars with de-reddened colors is presented as a thick solid line \citep{meyer97}. Cross symbols 
correspond to objects shown in Figure \ref{fig:2mass_lc}. The Galactic extinction is not considered here.
}
\label{fig:2mass_cc}
\end{figure}

Figure \ref{fig:2mass_cc} shows colors of the matched 2MASS objects with unblended, unsaturated, and accurate 
photometry which is described by read flag Rflg = 2, blend flag Bflg = 1, and contamination and 
confusion flag Cflg = 0 in all three bands of 2MASS data \citep{covey07}, and with the separation between 
the NSVS and 2MASS positions less than 1$''$. Most variable candidates have 
colors similar to those of normal stars in our Galaxy which are mainly $0 < J - H < 1$ \citep[e.g.,][]{finlator00,zoccali03}. Although 
quasars have distinctive colors in $J - H$ versus $H - K_{s}$ where most variable candidates are not found 
\citep[see][for discussions]{chiu07,kouzuma10}, some fraction of variable candidates might be quasars with colors around 
$J - H \sim 0.9$ and $H - K_{s} \sim 0.3$.

Late-type variable candidates such as red giant branch (RGB) and AGB stars can be identified more reliably with the 2MASS colors. 
As shown in Figure \ref{fig:2mass_cc}, the dominant NIR color of pulsating variable stars found in the Magellanic clouds 
\citep{ita04} is different from the major colors of our variable candidates and known quasars. In particular, $J - K_{s} 
\sim 1.4$ is the boundary between oxygen-rich and carbon stars \citep{nikolaev00,cole02,kiss03}. Therefore, 
our variable candidates with colors similar to those of known variables might be pulsating RGB and AGB stars \citep{ita04,kouzuma09}.

By using the NIR color -- color diagram given in Figure \ref{fig:2mass_cc}, we can also investigate whether our variable candidates 
include possible obscured young stars with or without disks such as T Tauri stars \citep{meyer97,tsujimoto02,ozawa05}. 
If reddening is significant in some of our variable candidates, their colors might be consistent with those of 
young stars.

\begin{figure}
\plotone{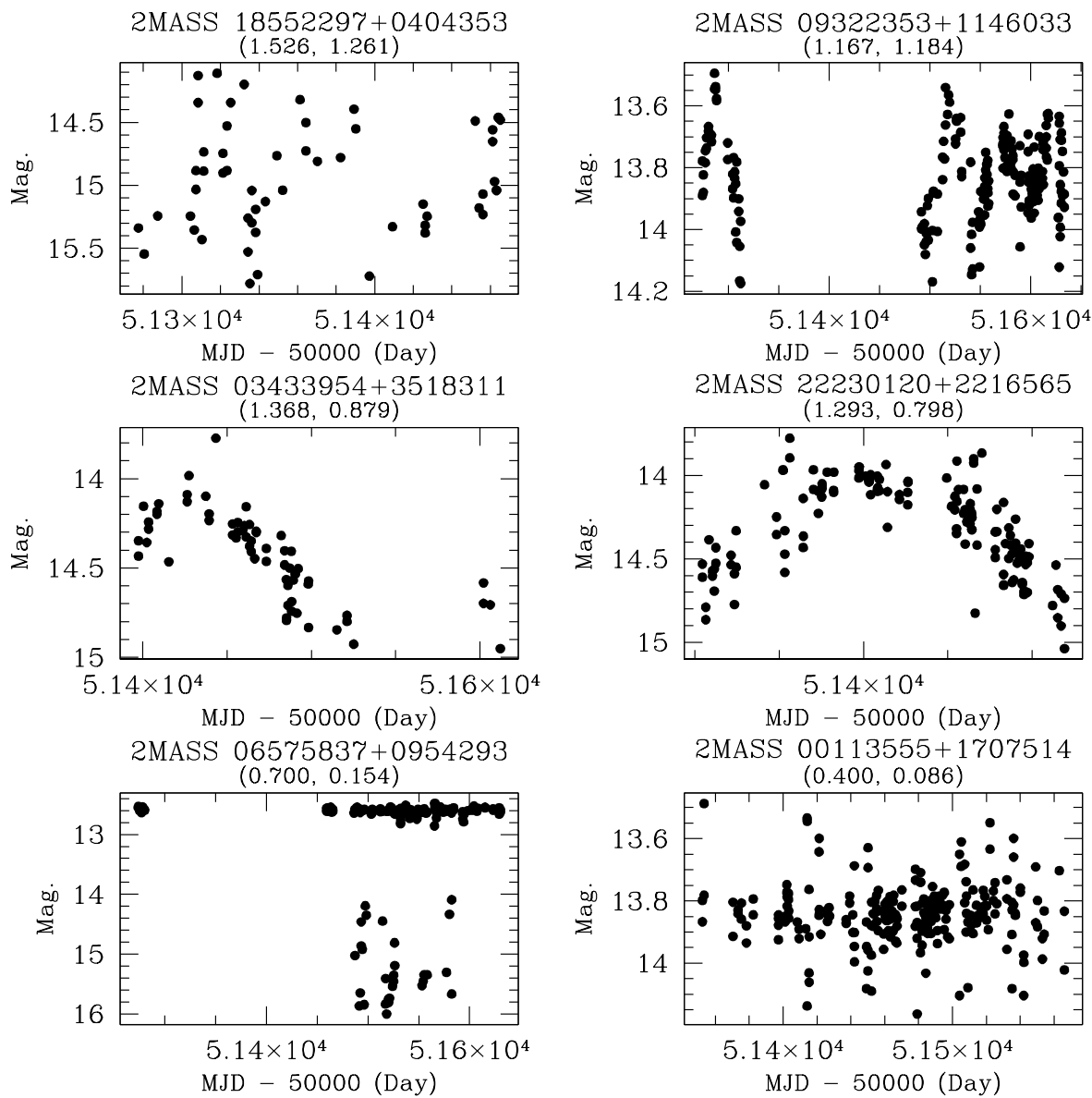}
\caption{Example light curves of the NSVS objects with the reliable 2MASS photometry. The 2MASS 
designations and ($J - H$, $H - K_{s}$) are given in the top of each panel. 2MASS 18552297+0404353 is 
also PDS 551 which is a Herbig Ae/Be candidate star \citep{vieira03}. 2MASS 09322353+1146033 corresponds to 
IRAS 09296+1159.
}
\label{fig:2mass_lc}
\end{figure}

Figure \ref{fig:2mass_lc} shows six example light curves of the NSVS objects which have corresponding 2MASS 
measurements. \object{2MASS 18552297+0404353} (NSVS 13924374) is a Herbig Ae/Be candidate star \citep{vieira03} with a 
different name PDS 551. The source \object{2MASS 09322353+1146033} (NSVS 10229563 and \object{IRAS 09296+1159}) is a post-AGB star \citep{blommaert93}. We find that 
its variation recorded in the NSVS light curve is regular with a period of about 46.88 days\footnote{
This period is found by using the tool provided in \url{http://www.astro.lsa.umich.edu/~msshin/science/code/MultiStep_Period/} 
\citep{shin04}.}. 
\object{2MASS 22230120+2216565} (NSVS 11767619) is confirmed as a carbon star in spectroscopic observations by \citet{mauron07}. 
But other objects presented in Figure \ref{fig:2mass_lc} have not been assigned a type.

We also match the NSVS coordinates of the variable candidates to the SDSS Data Release 7 with 6$''$ search radius. The 
five bands of the SDSS photometric systems have been commonly used to identify stellar and non-stellar sources 
by using their distinctive colors \citep[e.g.,][]{fan99,fukugita11}. Combining the 2MASS photometry with the SDSS photometry also helps 
us determine stellar source types precisely \citep[e.g.,][]{finlator00}. Among 438,087 variable candidates having 
corresponding SDSS photometric objects, 406,564 candidates have also corresponding 2MASS sources.

\begin{figure}
\plotone{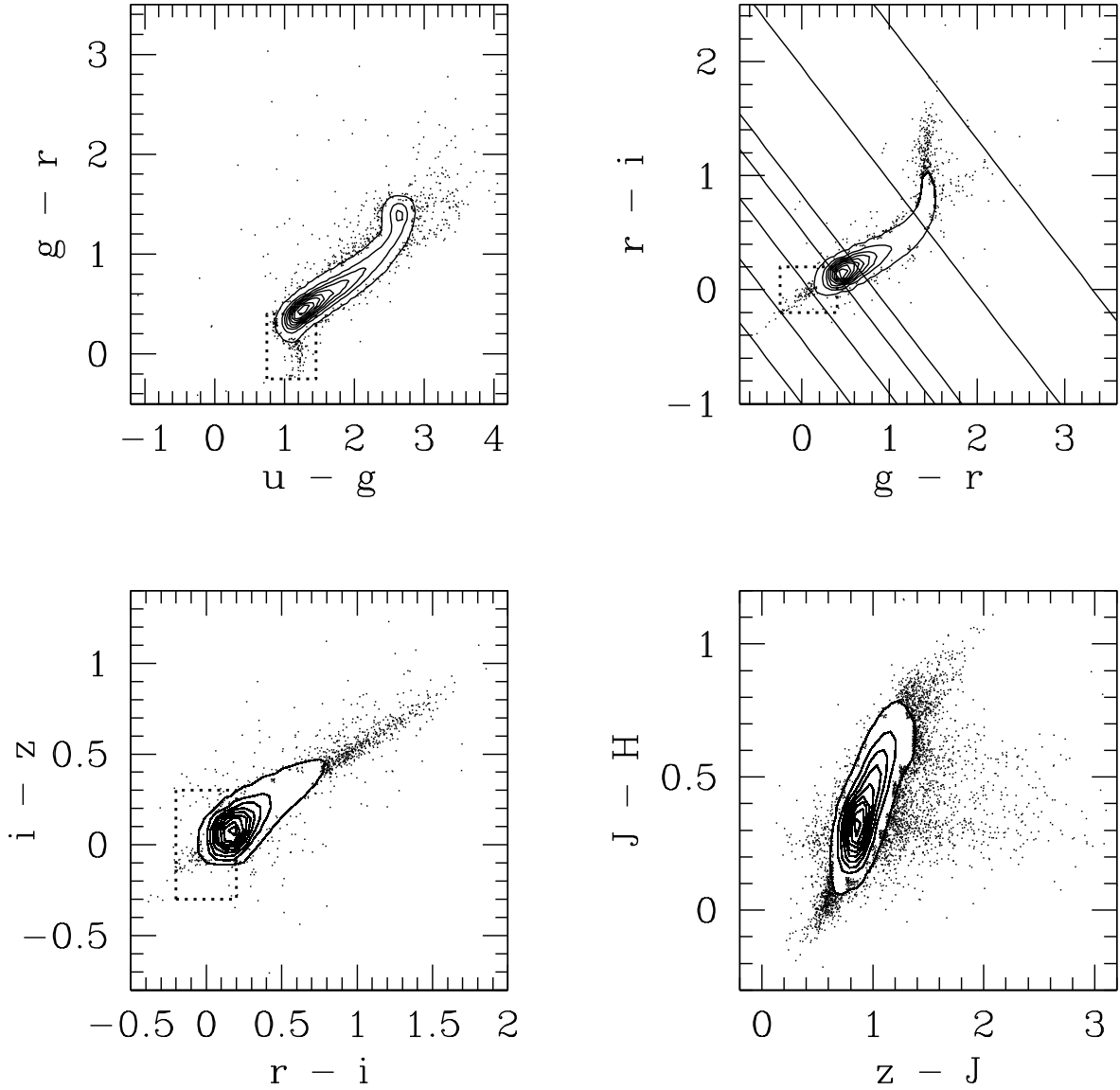}
\caption{SDSS color--color diagrams of the variable candidates. 
Boxes represent the ranges of single-epoch colors for RR Lyrae variable candidates from \citet{sesar10}. 
Solid lines in the panel of ($g - r$) and ($r - i$) colors represent ($g - i$) colors corresponding to 
spectral types O5, A0, F0, G0, K0, M0, and M5 from left to right, which are derived from synthesized stellar 
spectra \citep{covey07}. 
The Galactic extinction is not included here.
}
\label{fig:2mass_sdss_cc}
\end{figure}

Figure \ref{fig:2mass_sdss_cc} presents the distribution of colors for the variable candidates. We plot only good SDSS photometric 
data and clean 2MASS photometric data as we explained earlier. In particular, the matched objects with less than 1$''$ distance are 
shown in the plot. The good SDSS photometric data are defined as stellar (i.e., unresolved) objects without the SDSS photometric flags 
EDGE, BLENDED, PEAKCENTER, NOPROFILE, COSMIC\_RAY, SATURATED, NOTCHECKED, DEBLENDED\_AS\_MOVING, SATUR\_CENTER, 
INTERP\_CENTER, DEBLEND\_NOPEAK, and PSF\_FLUX\_INTERP \citep{stoughton02}. We check these flags in each SDSS band. Therefore, 
the number of objects shown in each panel of Figure \ref{fig:2mass_sdss_cc} varies for different color combinations. In all 
SDSS photometric data, we use PSF magnitudes. We also find SDSS objects with limiting magnitudes $m_{u} = 22.3$, 
$m_{g} = 23.3$, $m_{r} = 23.1$, $m_{i} = 22.3$, and $m_{z} = 20.8$.

The SDSS color -- color diagram can be used to pick out probable RR Lyrae variables which are 
pulsating horizontal branch stars \citep{gautschy96}. As suggested in the theoretical prediction of colors for RR Lyrae 
stars in the SDSS photometric system \citep{marconi06}, the following color ranges can be used to find RR Lyrae candidates 
\citep{sesar10}:
\begin{eqnarray}
0.75 < u - g < 1.45, \\
-0.25 < g - r < 0.4, \\
-0.2 < r - i < 0.2, \\
-0.3 < i - z < 0.3,
\end{eqnarray}
which are shown as boxes in Figure \ref{fig:2mass_sdss_cc}. A large number of variable candidates are identified 
as F-, G-, and K-type stars in the figure.

\begin{figure}
\plotone{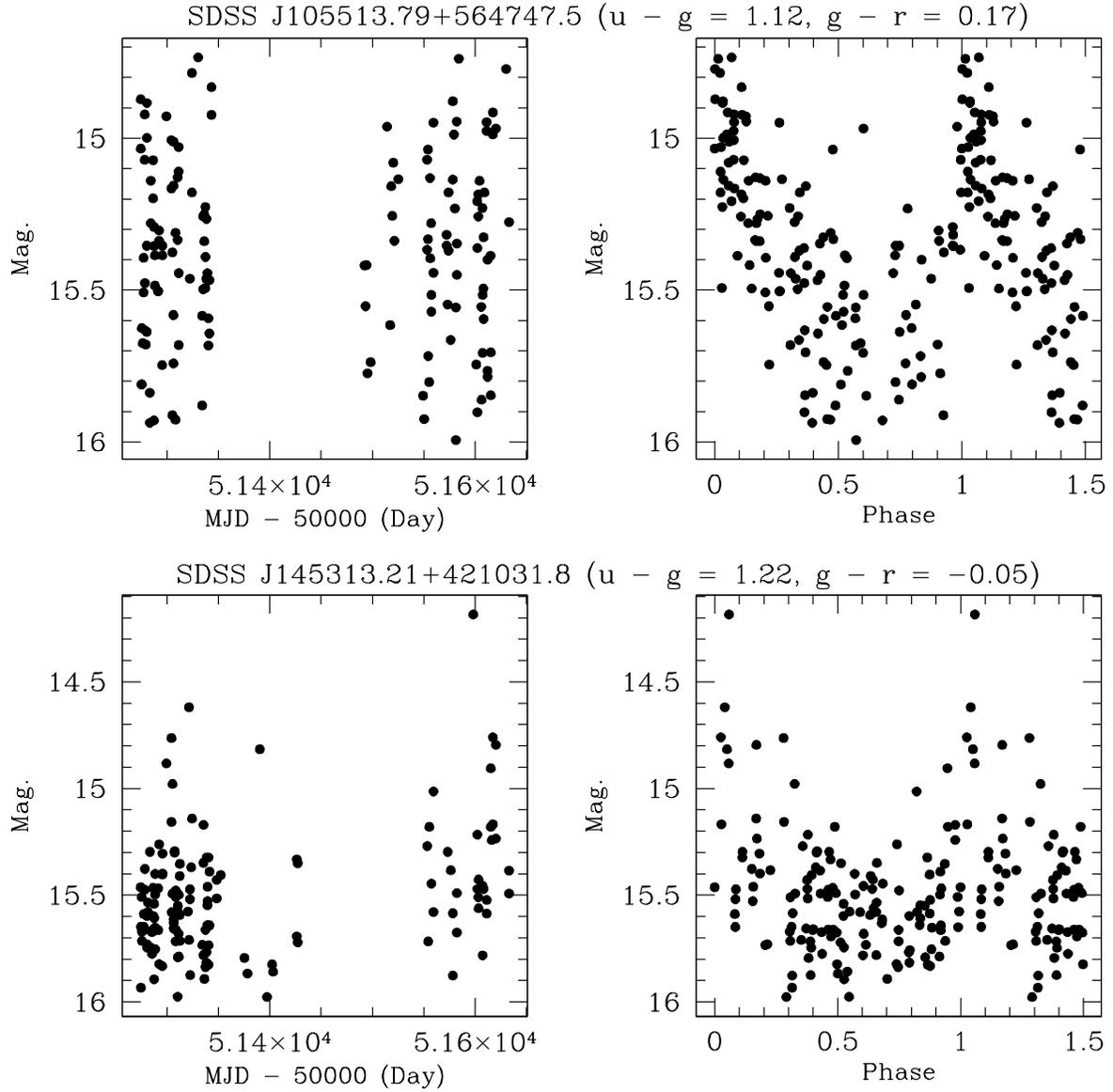}
\caption{Example light curves of the NSVS objects selected as RR Lyrae variable candidates with the SDSS 
spectroscopic data. The left column shows the raw NSVS light curves, while the right column presents 
the light curves folded with approximate periods of 0.541757 (top) and 0.489448 (bottom) 
days, respectively.
}
\label{fig:sdss_RR}
\end{figure}

Among the variable candidates selected by the RR Lyrae color cuts, we present example light curves of two objects, 
\object{SDSS J105513.79+564747.5} (NSVS 2594623) and \object{SDSS J145313.21+421031.8} (NSVS 5152328), 
which have been also observed in the SDSS spectroscopy, 
in Figure \ref{fig:sdss_RR}. Both objects are not classified as variable sources in the SIMBAD database. 
However, \object{SDSS J105513.79+564747.5} is found variable in GALEX observations \citep{welsh05} as included in 
the VSX catalog (see Appendix). We can estimate approximate periods of these two 
variables with the NSVS light curves as 0.541757 and 0.489448 days, respectively. These examples clearly show that 
a low cadence in the NSVS data is not high enough to derive complete light curves of these RR Lyrae variables, which have 
short periods \citep{sterken05}, except few NSVS objects with enough data \citep{kinemuchi06}. 
Therefore, further follow-up observations of interesting NSVS objects will be required 
to confirm their variability classes. \object{SDSS J105513.79+564747.5} was identified as a probable RR Lyrae by \citet{wheatley08}, 
although they could not retrieve a complete light curve from their {\it GALEX} observations. \object{SDSS J145313.21+421031.8} was 
also recognized as a blue horizontal branch star in the SDSS observation \citep{sirko04}. These examples prove that 
using colors of objects is complementary to variability analysis to identify object types.

\begin{figure}
\plotone{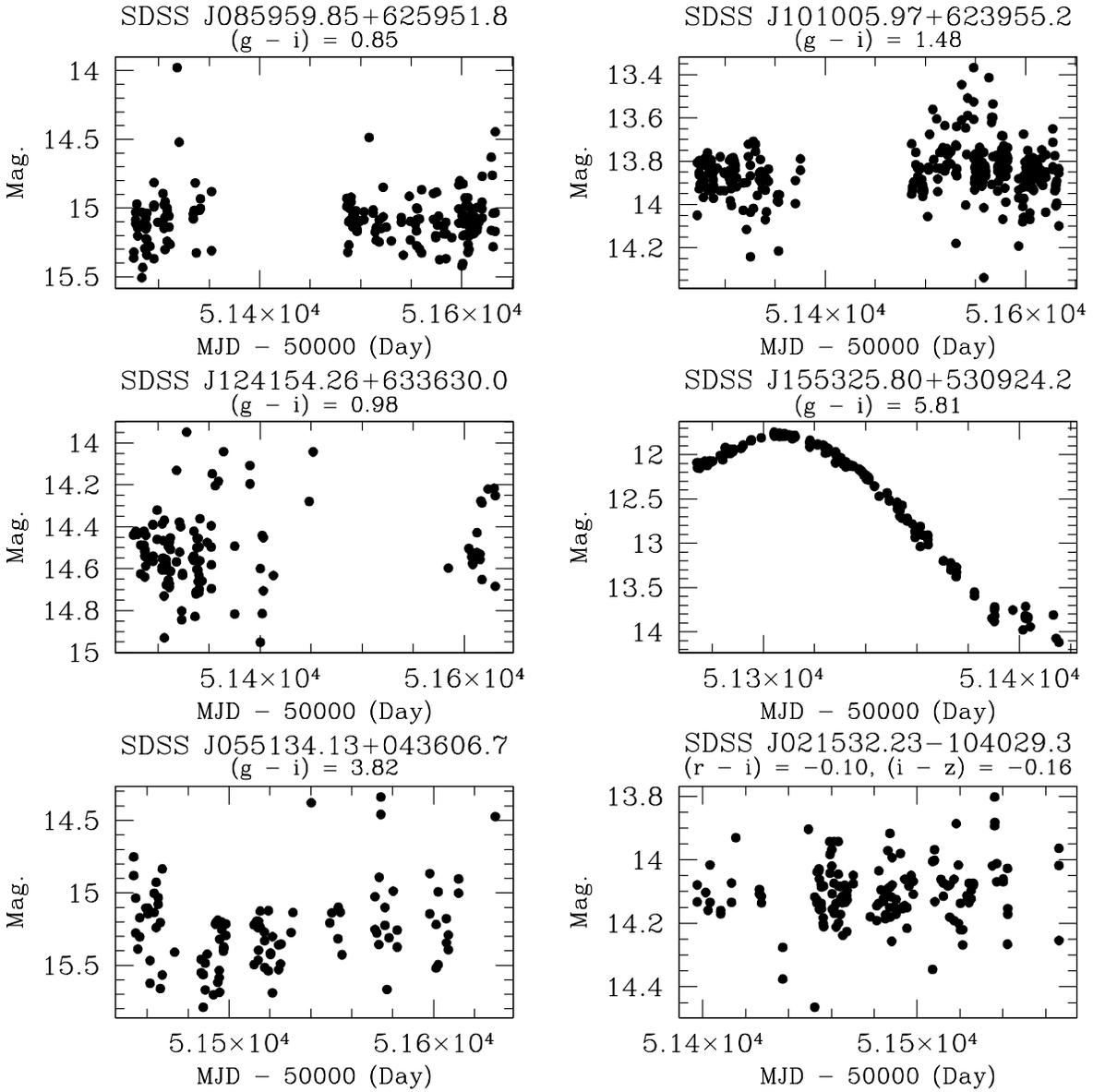}
\caption{Example light curves of variable candidates with the reliable SDSS and 2MASS photometric measurements. 
Top two objects are included in all color--color diagrams shown in Figure \ref{fig:2mass_sdss_cc}. 
SDSS J155325.80+530924.2 is not classified as a known variable star in the SIMBAD database. However, it is a spectroscopically 
confirmed giant star with another designation 2MASS J15532581+5309244 \citep{cruz03}, and is found as a variable star in \citet{wozniak04b}. 
SDSS J021532.23-104029.3 is identified as a field horizontal branch star BPS CS 22175-0003 
by \citet{wilhelm99}.
}
\label{fig:sdss_lc}
\end{figure}

Six other examples of variable candidates are presented in Figure \ref{fig:sdss_lc} where NSVS objects have 
corresponding reliable SDSS and 2MASS photometric data. Except for \object{SDSS J021532.23-104029.3}, these objects 
have ($g - i$) colors without the SDSS bad photometric flags. Spectral types of stars can be 
described approximately by ($g - i$) colors where B0, A0, F0, G0, K0, and M0 correspond to 
($g - i$) $\sim$ $-0.94$, $-0.44$, 0.09, 0.52, 0.83, and 1.95, respectively 
\citep{covey07}. Most variable candidates are close to G5 as shown in Figure \ref{fig:2mass_sdss_cc}. 
Although ($g - i$) is generally a good proxy of spectral types, \object{SDSS J155325.80+530924.2} (NSVS 5206326) was 
already confirmed spectroscopically as M8 III star for its ($g - i$) = 5.81. We also note that this object is 
also included in the VSX catalog as a variable star (see Appendix). These examples reassert 
that the low cadence in the NSVS data does not guarantee a certain classification of variable objects, requiring 
further follow-up observations with different cadences.

\subsection{SDSS and {\it GALEX} photometry}

Hot stellar objects such as white dwarfs and massive main-sequence stars are generally not detected in 
NIR, but they can be more easily recognized over UV wavelength ranges which cover most stellar flux. In the 
{\it GALEX} GR4/5\footnote{\url{http://galex.stsci.edu}}, we find objects matching our NSVS variable candidates' 
coordinates within 6$\arcsec$. When multiple {\it GALEX} objects are matched to a single NSVS object, we choose the nearest 
{\it GALEX} object as the best match. A total of 739,625 variable candidates, 
i.e., about 40\% of the candidates, correspond to {\it GALEX} photometric objects. 
A total of 286,185 candidates have corresponding SDSS objects too.

\begin{figure}
\plotone{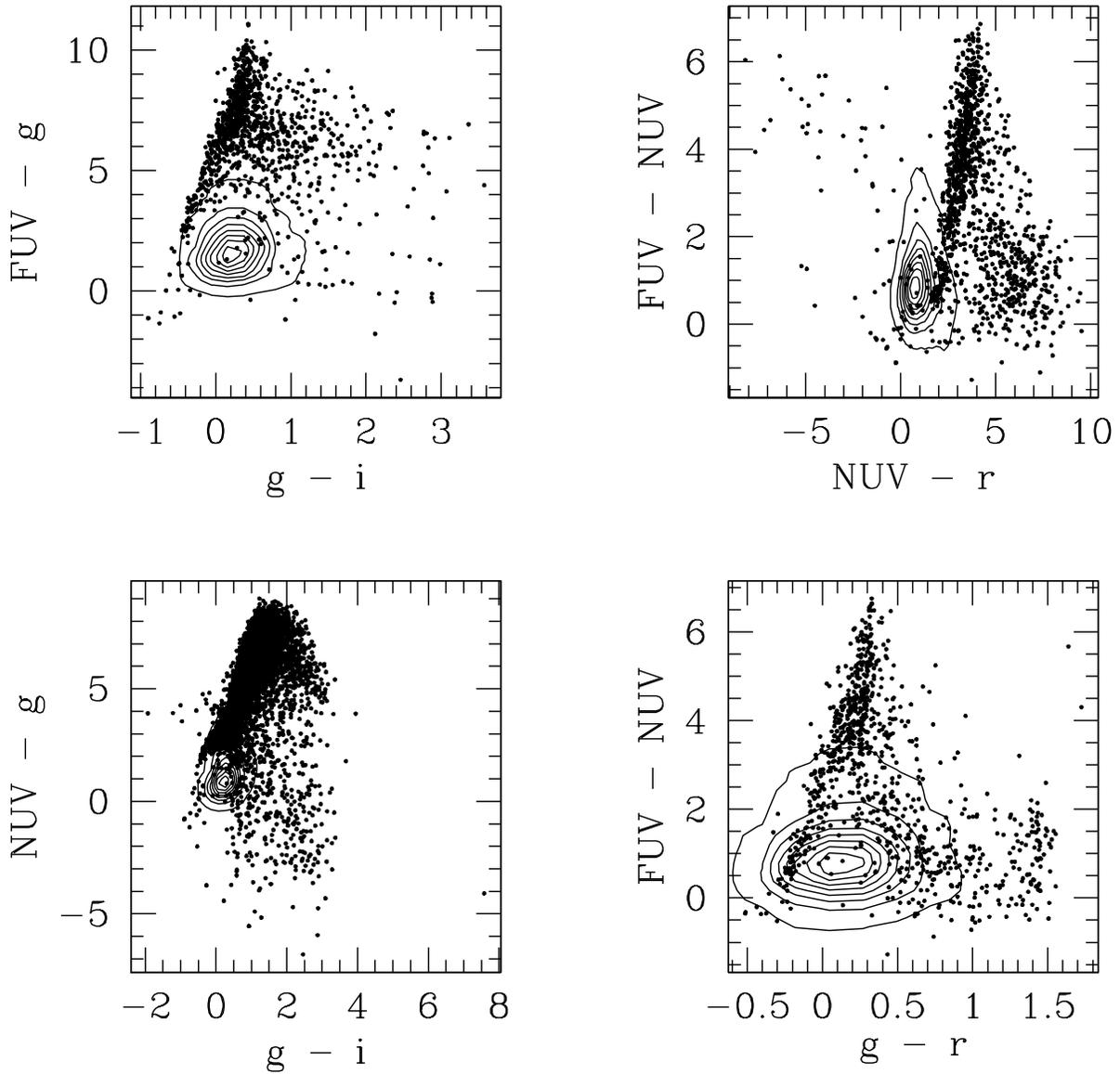}
\caption{Color-color diagrams of variable candidates with the SDSS and {\it GALEX} photometric data. The plots 
show variable candidates matching the SDSS objects within 1$''$. Contours correspond to the color 
distributions of quasars which are detected in both SDSS and {\it GALEX} \citep{trammell07}.
}
\label{fig:galex_sdss_cc}
\end{figure}

In Figure \ref{fig:galex_sdss_cc}, we present colors of variable candidates with reliable {\it GALEX} and SDSS photometric 
measurements. In addition to following the same conditions for the reliable SDSS photometric measurements as in Section 
\ref{sec:2mass_sdss}, we require that the {\it GALEX} objects should have 
the distance from the center of the {\it GALEX} field of view $< 0.^{\circ}6$, and both FUV and NUV magnitudes $< 25$ 
\citep{agueros05,maxted09}. The FWHM angular resolution is about 6$''$ in the NUV channel \citep{morrissey05}. 
Considering the combined effects of poor spatial resolution in the NSVS and {\it GALEX} data, the spatial association among 
the NSVS, SDSS, and {\it GALEX} objects needs a careful check when people select interesting NSVS objects with 
the corresponding SDSS and {\it GALEX} objects together in our database.

Colors of most variable candidates presented in Figure \ref{fig:galex_sdss_cc} are consistent with the expected colors 
of normal stars \citep{seibert05,bianchi07,bianchi09}, considering the Galactic extinction that makes overall colors red. The color 
distribution of known quasars \citep{trammell07} is well separated from that of stars in the diagram of (${\rm FUV} - g$) and ($g - i$). 
Because our variable candidates are bright objects, all of them might not be quasars but stars even though some objects seem 
to have quasar-like colors. Hot white dwarf 
candidates can be selected with the color cut of (${\rm FUV - NUV}$) $< 0$ and ($g - r$) $< -0.2$ \citep{agueros05}. 
But we warn that the multiple matches of the NSVS coordinates to both SDSS and {\it GALEX} catalogs have worse 
precision than a single match to either SDSS or {\it GALEX} catalogs. Therefore, the color combining both 
SDSS and {\it GALEX} photometric data might not be reliable when the objects are faint or close to neighboring objects 
in the SDSS and {\it GALEX} catalogs. Since the precision of the SDSS objects' coordinates is much better than those of the {\it GALEX} catalogs, 
the SDSS colors are more reliable than the {\it GALEX} colors in the color -- color diagram combining both catalogs.

\begin{figure}
\plotone{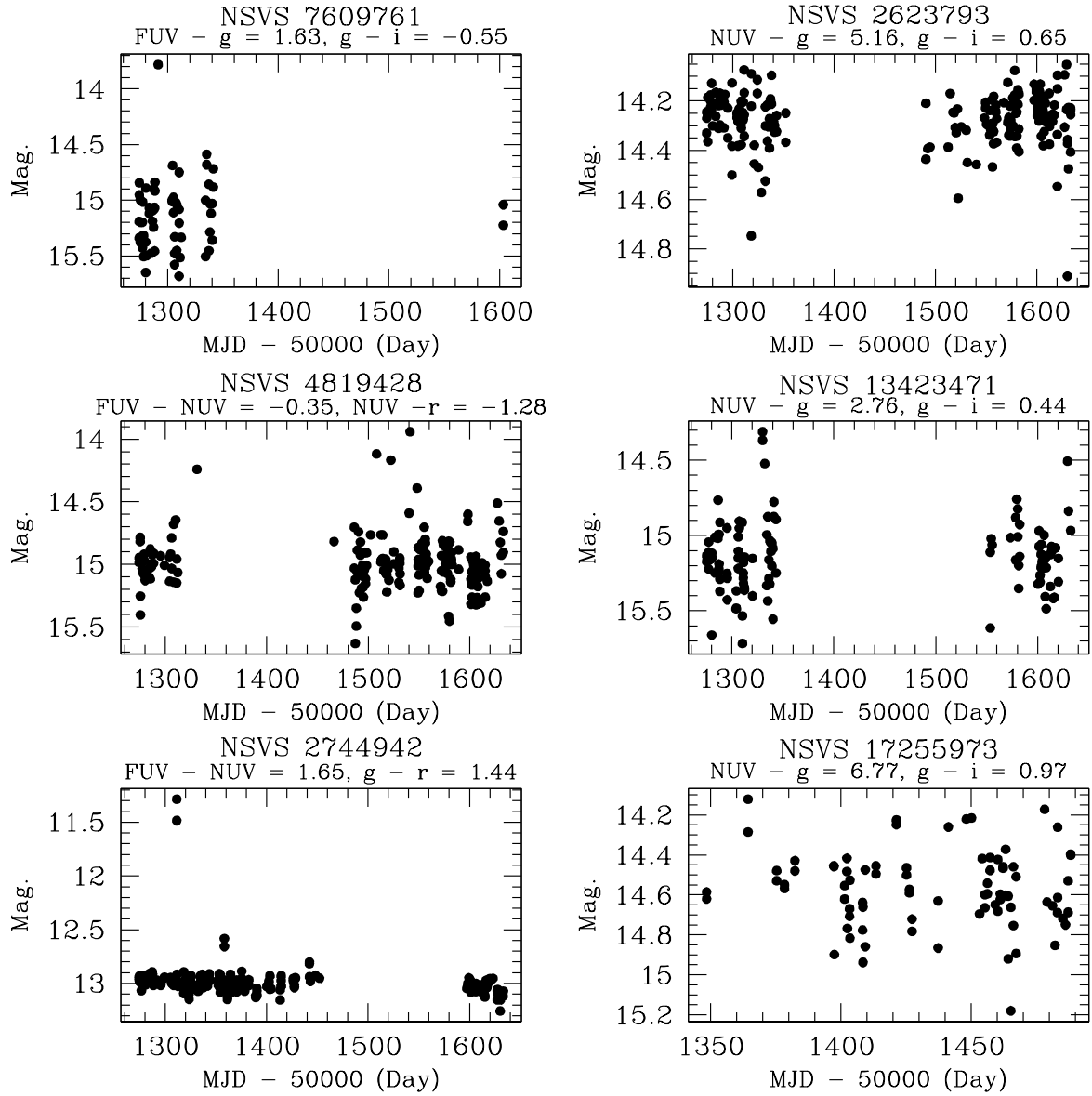}
\caption{Example light curves of variable candidates with the reliable {\it GALEX} and SDSS photometry. 
Three objects (left) have further identification in SIMBAD, while other three objects 
(right) are selected in the order of increasing (${\rm NUV} - g$) color from top to bottom.
}
\label{fig:galex_sdss_lc}
\end{figure}

Among the NSVS objects with the corresponding SDSS and {\it GALEX} objects, several SIMBAD objects are found with 
further information about their properties. For example, NSVS 7609761 corresponds to \object{SDSS J115800.38+295731.4} and 
\object{GALEX J115800.4+295731} with (${\rm FUV - NUV}$) $= 1.63$ and ($g - i$) $= -0.55$. This object is included in \citet{brown08} 
as \object{CHSS 835} which is a star with a spectral type of B8. 
In the diagram of (${\rm FUV - NUV}$) and (${\rm NUV} - r$), NSVS 4819428 is known as a spectral type B subdwarf \object{FBS 0839+399} 
\citep{wegner85,mickaelian08} with (${\rm FUV - NUV}$) $= -0.35$ and (${\rm NUV} - r$) $= -1.28$. 
The light curves of these two 
objects given in Figure \ref{fig:galex_sdss_lc} do not show distinctive features 
due to the poor sampling rate in the NSVS data. Flare-like variation 
can be presumed from the light curve of NSVS 2744942 which is recognized as an active M dwarf, corresponding to 
an X-ray object \object{RX J1447.2+5701} \citep{mochnacki02}. 
The NSVS light curve of this object shows about 1.5 mag variation even with the poor sampling rate.

Figure \ref{fig:galex_sdss_lc} also shows the light curves of the three variable candidates which have reliable data 
of the SDSS $g-$ and $i$-band photometry as well as the {\it GALEX} NUV measurement. None of them have any different identification 
in SIMBAD database. The light variation seems real, but the low cadence in the NSVS data does not produce complete light curves 
with distinctive types.

\section{Summary and discussion}\label{sec:last}

A new systematic investigation of variable candidates in the NSVS data was presented with a clustering method for time-series data. 
Assuming that the dominant fraction of light curves represents non-variable objects, our method finds clusters of light curves with 
their eight dimensional features, and then finds how many light curves are included in each cluster. 
When choosing as variable candidate objects which are not included in the top four large clusters and which have $D_{M} >$ 99\% cut 
from the largest cluster, the total number of new variable candidate light curves is 1,824,123 in our entire sample of the NSVS data.

The cross-correlation with {\it IRAS}, AKARI, 2MASS, SDSS, and {\it GALEX} catalogs helps us to identify interesting objects with specific spectral 
types or variability classes \citep{eyer08}. In particular, variable stars over the instability strip can be selected easily from their specific colors 
(e.g., see Figure \ref{fig:sdss_RR}). 
We also show examples of long-period variables which can be selected from their {\it IRAS}, AKARI, or 2MASS colors 
(e.g., see Figures \ref{fig:iras_lc}, \ref{fig:akari_lc}, and \ref{fig:2mass_lc}).

Our analysis is presented online with the information on cross-correlations with other catalogs. Because the sampling pattern in 
the NSVS data is not good enough to identify detailed structures of different variability types, follow-up observations of variable 
candidates will be necessary to understand these variable sources by taking more photometric data points and improving 
the sampling rates. Moreover, some variable candidates such as {\it IRAS} sources might be new maser sources which are interesting 
objects in the radio region.

Our approach of detecting variable candidates in the NSVS data is supplementary to previous methods of 
finding variable candidates. We do not claim that this method is the best way in all cases. Definitely, 
if observation systems, including instruments, environments, and data reduction procedures, are well 
known prior or are well controlled, supervised methods can be superior than unsupervised methods like 
our approach because supervised methods can simulate observing systems with known variable and 
non-variables sources to find the best separation between variable and non-variable sources. This separation 
can be applied to detect new variable candidates in the test data. Therefore, when the data properties, including 
all kinds of systematic patterns and real variability patterns, are understood and modeled well, detecting 
variable sources becomes a {\it classification} problem instead of a {\it clustering} problem.

Our analysis results can be used with many different methods of selecting variable candidates. In this paper, our conservative 
selection method is avoiding the top four large clusters and objects with $D_{M} <$ 99\%. However, when people 
are interested in infrared variable sources, they can choose as variable candidate objects corresponding to {\it IRAS} objects with  $D_{M} >$ 90\%. 
If known variable 
objects can produce clusters with reasonable sizes, finding clusters with many known variable objects can be an 
efficient way to find variable candidates. Unfortunately, this approach is not feasible now 
because the number of known variable objects is too small in each NSVS observation field.

Several variability surveys cover the same apparent magnitude ranges as the NSVS does. A large fraction of sky has been 
already observed in the All Sky Automated Survey \citep[ASAS;][]{pojmanski97} and SuperWASP \citep{street03}. Our approach of 
variability detection can be applied to those data sets too. In addition, objects included in both the NSVS and 
others can be combined to extend the span of time-series data or to complement different sampling patterns. 
We plan to update the online database with these additional data sets in the future. Furthermore, because almost all objects included in our 
analysis will be monitored by the {\it GAIA} mission for five years \citep{gaia}, our analysis will be combined with the future 
time-series data and astrometric/kinematic information.

Our method can also be improved to catch much broader types of variable objects and objects with weak variability 
signals. For this purpose, it is important to include various features of light curves as we emphasized in Section \ref{sec:gmm}. 
In particular, the usage of AoVM as one feature of light curves is strongly limited because a complete 
form of a periodogram has more information of light curves. Therefore, it must be useful to develop 
new features describing periodograms more completely if the new features can be 
estimated in computationally cheap ways. Moreover, Stetson's $I$ index \cite{stetson96} can be included in 
our method if data include multi-band light curves. For instance, $J$ and $K$ can be derived in each band, while 
$I$ is estimated with multi-band light curves. Our usage of $J$ can also be changed to use all pairs of data points 
instead of using sequential pairs.

\acknowledgments

We thank Bernie Shao and Przemek Wozniak for helping us access the {\it GALEX} data and the NSVS light curves, 
respectively. We also thank Michael Sekora, Charles Cowley, Mark Reynolds, and Stefan Kraus 
for useful discussions and careful reading. We are grateful to the anonymous referee for comments which 
improved this manuscript. Y.I.B. acknowledges the support of National Research 
Foundation of Korea through Grant 2011-0030875.

This publication makes use of the data from the Northern Sky Variability Survey created 
jointly by the Los Alamos National Laboratory and University of Michigan. The NSVS was funded 
by the Department of Energy, the National Aeronautics and Space Administration, and 
the National Science Foundation. 
This research has made use of the SIMBAD database,
operated at CDS, Strasbourg, France. 
This research is based on observations with AKARI, a JAXA project with the participation of ESA. 
This publication also makes use of data products from the Two Micron All Sky Survey, 
which is a joint project of the University of Massachusetts and the 
Infrared Processing and Analysis Center/California Institute of Technology, 
funded by the National Aeronautics and Space Administration and the National Science Foundation. 
Funding for the SDSS and SDSS-II has been provided by the Alfred P. Sloan Foundation, 
the Participating Institutions, the National Science Foundation, the U.S. Department of 
Energy, the National Aeronautics and Space Administration, the Japanese Monbukagakusho, 
the Max Planck Society, and the Higher Education Funding Council for England. The SDSS 
Web Site is http://www.sdss.org/. 
The SDSS is managed by the Astrophysical Research Consortium for the Participating Institutions. 
The Participating Institutions are the American Museum of Natural History, Astrophysical Institute 
Potsdam, University of Basel, University of Cambridge, Case Western Reserve University, University 
of Chicago, Drexel University, Fermilab, the Institute for Advanced Study, the Japan Participation 
Group, Johns Hopkins University, the Joint Institute for Nuclear Astrophysics, the Kavli Institute for 
Particle Astrophysics and Cosmology, the Korean Scientist Group, the Chinese Academy of Sciences (LAMOST), 
Los Alamos National Laboratory, the Max-Planck-Institute for Astronomy (MPIA), the Max-Planck-Institute 
for Astrophysics (MPA), New Mexico State University, Ohio State University, University of Pittsburgh, 
University of Portsmouth, Princeton University, the United States Naval Observatory, and the University of Washington.
GALEX (Galaxy Evolution Explorer) is a NASA Small Explorer, launched in April 2003. We gratefully 
acknowledge NASA's support for construction, operation, and science analysis for the GALEX mission, 
developed in cooperation with the Centre National d'Etudes Spatiales (CNES) of France and the 
Korean Ministry of Science and Technology. 

\appendix

\section{Known or suspected variables found in the VSX}

The SIMBAD database is commonly used to identify known objects. We also use the database to find all known 
variable objects and extragalactic objects as shown in Section \ref{sec:simbad}. However, the SIMBAD database 
is not as complete as other catalogs of variable stars. In particular, the AAVSO International Variable Star 
Index (VSX)\citep{watson06} catalog is frequently updated with new reports of variable stars.

We check how many variable candidates selected by our conservative selection are not classified as variable 
sources in the SIMBAD database, but are included in the VSX catalog. Here, we use the catalog released online on Nov. 15, 2009, 
including 178,599 stars. Among 1,824,123 variable candidates chosen by our conservative selection and not included in 
the SIMBAD database, we find that 41,019 objects are included as variable stars in the VSX catalog. These known variable stars 
in the catalog are largely included with references to the ASAS observations (9,999 objects) \citep{pojmanski97}, and to the NSVS 
observations (15,958 objects) that we also use here. A total of 1,147 objects are matched to suspected variable stars in the catalog. 
Interestingly, a total of 1,106 suspected variables are included with references to the NSVS 
light curves. Therefore, at least 1,783,104 variable candidates are newly selected in our method.
In the online database, we provide links to the most recent VSX catalog for the NSVS light curves which we examine.


\clearpage

\clearpage

\clearpage

\clearpage

\clearpage

\clearpage

\begin{figure}
\plotone{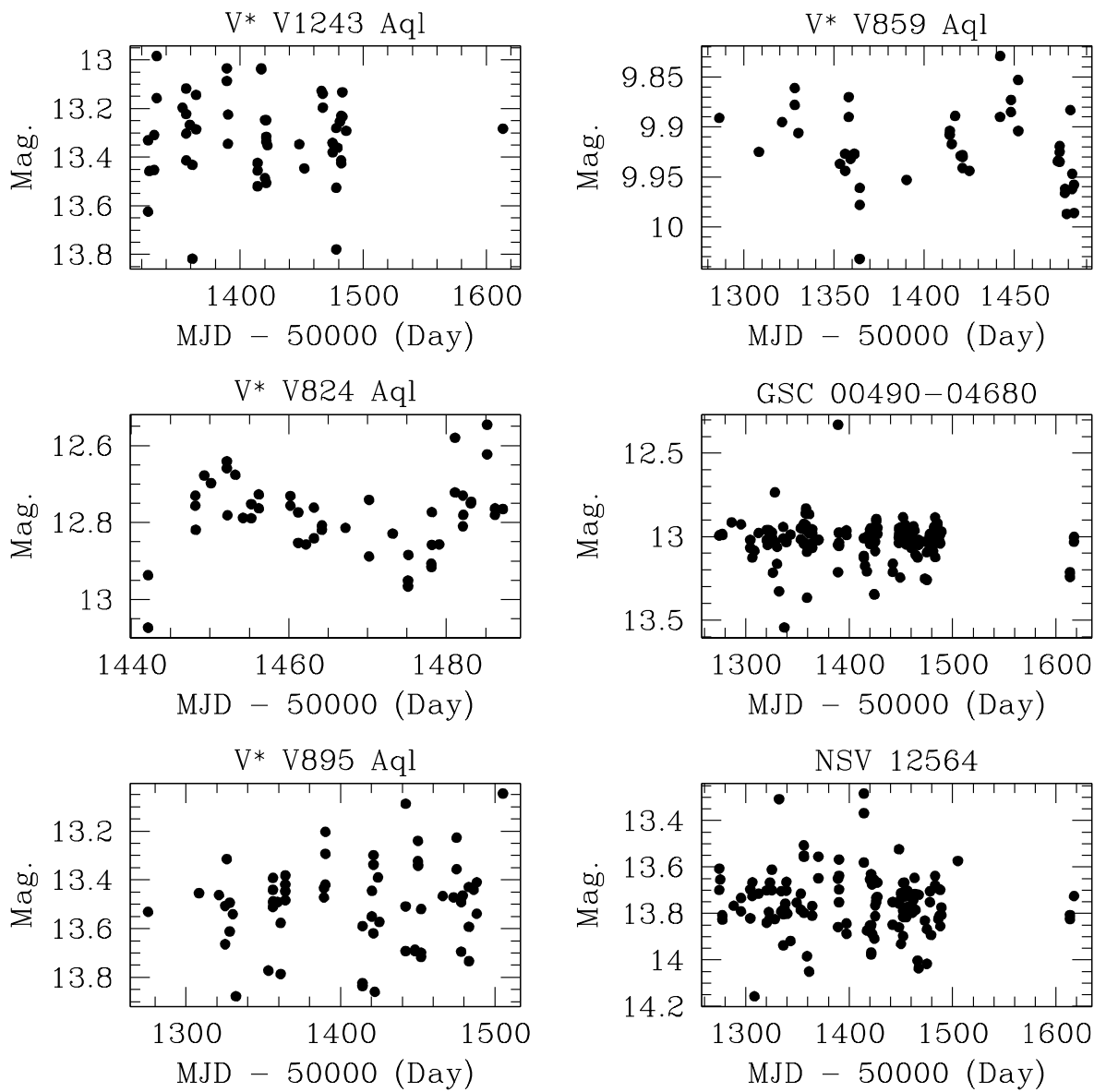}
\caption{Example light curves of known or suspected variable objects which are not selected by our conservative variability detection. 
The SIMBAD names of the objects are given in the top of each panel.}
\label{fig:missing_exam}
\end{figure}
\clearpage

\clearpage

\clearpage

\clearpage

\clearpage

\clearpage

\clearpage

\clearpage

\clearpage

\clearpage

\clearpage

\clearpage

\clearpage

\clearpage


\begin{deluxetable}{lc}
\rotate
\tablecaption{Variability indices.\label{tab:var}}
\tablehead{\colhead{Index} & \colhead{Definition}}
\startdata
$\frac{\sigma}{\mu}$ & $\frac{\sqrt{ \sum_{n=1}^{N}(x_{n} ~-~ \mu)^2 / (N ~-~ 1) }}{\sum_{n=1}^{N}x_{n} / N}$ \\ \\
$\gamma_{1}$ & $\frac{\sqrt{ N (N ~-~ 1) }}{N ~-~ 2} \frac{\sum_{n=1}^{N}(x_{n} ~-~ \mu)^3 / N}{\sqrt[3]{\sum_{n=1}^{N}(x_{n} ~-~ \mu)^2 / N}}$ \\ \\
$\gamma_{2}$ & $\frac{N ~-~ 1}{(N ~-~ 2)(N ~-~ 3)} 
\left\{ {(N ~+~ 1) \left(\frac{\sum_{n=1}^{N}(x_{n} ~-~ \mu)^4 / N}{(\sum_{n=1}^{N}(x_{n} ~-~ \mu)^2 / N)^{2}} ~-~ 3\right) ~+~ 6} \right\}$ \\ \\
Con & $
1 ~+~ \frac{1}{N ~-~ 2} \sum_{n=1}^{N-2}
\left\{
\begin{array}{lc}
1 & \rm{if} ~ (x_{n} - \mu_{0}) > 2\sigma ~ \rm{and} ~ (x_{n+1} - \mu_{0}) > 2\sigma ~ \rm{and} ~ (x_{n+2} - \mu_{0}) > 2\sigma \\
1 & \rm{if} ~ (x_{n} - \mu_{0}) < -2\sigma ~ \rm{and} ~ (x_{n+1} - \mu_{0}) < -2\sigma ~ \rm{and} ~ (x_{n+2} - \mu_{0}) < -2\sigma \\
0 & \rm{otherwise}
\end{array}
\right.
$ \\ \\
$\eta$ & $\frac{\sum_{n=1}^{N-1}(x_{n+1} ~-~ x_{n})^2 / (N ~-~ 1)}{\sigma^{2}}$ \\ \\
$J$ & $\sum_{n=1}^{N-1}{\rm sign}(\delta_{n}\delta_{n+1})\sqrt{\vert\delta_{n}\delta_{n+1}\vert}$ \\ \\
$K$ & $\frac{{1/N}\sum_{n=1}^{N}\vert\delta_{n}\vert}{\sqrt{1/N \sum_{n=1}^{N}\delta_{n}^{2}}}$ \\ \\
AoVM & The maximum value of the analysis of variance (ANOVA) statistic \citep{schwarzenberg96} \\
\enddata{}
\tablecomments{$\sigma$, $\mu$, $\gamma_{1}$, $\gamma_{2}$, and $\mu_{0}$ are standard deviation, average, skewness, 
kurtosis, and median of $N$ magnitudes $x_{n}$ in each light curve, respectively. $\delta_{n}$ is 
$\sqrt{N/(N-1)} (x_{n} - \mu)/e_{n}$ where $e_{n}$ is a photometric error for each data point. 
${\rm sign}(\delta_{n}\delta_{n+1})$ is the sign of $\delta_{n}\delta_{n+1}$.}
\end{deluxetable}


\begin{thebibliography}{}
\bibitem[Abazajian et al.(2009)]{sdss} Abazajian K.~N., et al., 2009, ApJS, 182, 543 
\bibitem[Ag{\"u}eros et al.(2005)]{agueros05} Ag{\"u}eros M.~A., et al., 2005, AJ, 130, 1022 
\bibitem[Alard \& Lupton(1998)]{alard98} Alard C., Lupton R.~H., 1998, ApJ, 503, 325 
\bibitem[Becker et al.(2004)]{becker04} Becker A.~C., et al., 2004, ApJ, 611, 418 
\bibitem[Bernhard \& Lloyd(2000)]{bernhard00} Bernhard K., Lloyd C., 2000, IBVS, 4920, 1 
\bibitem[Bianchi \& The GALEX Team(1999)]{bianchi99} Bianchi L., The GALEX Team, 1999, Memorie della Societ{\'a} Astronomica Italiana, 70, 365 
\bibitem[Bianchi et al.(2007)]{bianchi07} Bianchi L., et al., 2007, ApJS, 173, 659 
\bibitem[Bianchi et al.(2009)]{bianchi09} Bianchi L., Hutchings J.~B., Efremova B., Herald J.~E., Bressan A., Martin C., 2009, AJ, 137, 3761 
\bibitem[Blommaert, van der Veen, \& Habing(1993)]{blommaert93} Blommaert J.~A.~D.~L., van der Veen W.~E.~C.~J., Habing H.~J., 1993, A\&A, 267, 39 
\bibitem[Bono, Trevese, \& Turatto(2003)]{bono03} Bono G., Trevese D., Turatto M., 2003, MmSAI, 74, 1004 
\bibitem[Brown et al.(2008)]{brown08} Brown W.~R., Beers T.~C., Wilhelm R., Allende Prieto C., Geller M.~J., Kenyon S.~J., Kurtz M.~J., 2008, AJ, 135, 564 
\bibitem[Cacciari(2009)]{gaia} Cacciari, C.\ 2009, Memorie della Societ\`{a} Astronomica Italiana, 80, 97
\bibitem[Chen, Morris, \& Martin(2006)]{chen06} Chen  T., Morris  J., Martin  E., 2006, J. Roy. Stat. Soc. Ser. C, 55, 699
\bibitem[Chiu et al.(2007)]{chiu07} Chiu K., Richards G.~T., Hewett P.~C., Maddox N., 2007, MNRAS, 375, 1180 
\bibitem[Clegg(1980)]{clegg80} Clegg P.~E., 1980, PhyS, 21, 678 
\bibitem[Cole \& Weinberg(2002)]{cole02} Cole A.~A., Weinberg M.~D., 2002, ApJ, 574, L43 
\bibitem[Corwin et al.(2006)]{corwin06} Corwin T.~M., Sumerel A.~N., Pritzl B.~J., Smith H.~A., Catelan M., Sweigart A.~V., Stetson P.~B., 2006, AJ, 132, 1014 
\bibitem[Covey et al.(2007)]{covey07} Covey K.~R., et al., 2007, AJ, 134, 2398 
\bibitem[Coyne \& MacConnell(1983)]{coyne83} Coyne G.~V., MacConnell D.~J., 1983, VatOP, 2, 73 
\bibitem[Cruz et al.(2003)]{cruz03} Cruz K.~L., Reid I.~N., Liebert J., Kirkpatrick J.~D., Lowrance P.~J., 2003, AJ, 126, 2421 
\bibitem[Davies \& Bouldin(1979)]{davies79} Davies D.~L., Bouldin D.~W., 1979, IEEE Trans. on Pattern Analysis and Machine Intelligence, 1, 224
\bibitem[Dimitrov(2009)]{dimitrov09} Dimitrov D., 2009, Bulgarian Astronomical Journal, 12, 49 
\bibitem[Djorgovski et al.(2001)]{djorgovski01} Djorgovski S.~G., Mahabal A.~A., Brunner R.~J., Gal R.~R., Castro S., de Carvalho R.~R., Odewahn S.~C., 2001, ASPC, 225, 52 
\bibitem[Dommanget \& Nys(1994)]{dommanget94} Dommanget J., Nys O., 1994, Catalog of the components of double and multiple stars, Com. de l'Observ. Royal de Belgique, 115, 1 
\bibitem[Eyer(2006)]{eyer06} Eyer L., 2006, ASPC, 349, 15 
\bibitem[Eyer \& Mowlavi(2008)]{eyer08} Eyer L., Mowlavi N., 2008, JPhCS, 118, 012010 
\bibitem[Fan(1999)]{fan99} Fan X., 1999, AJ, 117, 2528 
\bibitem[Finlator et al.(2000)]{finlator00} Finlator K., et al., 2000, AJ, 120, 2615 
\bibitem[Fukugita et al.(2011)]{fukugita11} Fukugita, M., Yasuda, N., Doi, M., Gunn, J.~E., \& York, D.~G.\ 2011, \aj, 141, 47 
\bibitem[Gautschy \& Saio(1996)]{gautschy96} Gautschy A., Saio H., 1996, ARA\&A, 34, 551 
\bibitem[G{\"o}ssl \& Riffeser(2002)]{gossl02} G{\"o}ssl C.~A., Riffeser A., 2002, A\&A, 381, 1095 
\bibitem[Hartman et al.(2004)]{hartman04} Hartman, J.~D., Bakos, G., Stanek, K.~Z., \& Noyes, R.~W.\ 2004, AJ, 128, 1761 
\bibitem[Helou \& Walker(1988)]{helou88} Helou G., Walker D.~W., Infrared Astronomical Satellite (IRAS) Catalogs and Atlases, Vol. 7. NASA, Washington, DC
\bibitem[Hoffman et al.(2008)]{hoffman08} Hoffman D.~I., Harrison T.~E., Coughlin J.~L., McNamara B.~J., Holtzman J.~A., Taylor G.~E., Vestrand W.~T., 2008, AJ, 136, 1067 
\bibitem[Hoffman, Harrison, \& McNamara(2009)]{hoffman09} Hoffman D.~I., Harrison T.~E., McNamara B.~J., 2009, AJ, 138, 466 
\bibitem[Ishihara et al.(2010a)]{akari_irc} Ishihara, D., et al.\ 2010, A\&A, 514, A1 
\bibitem[Ishihara et al.(2010b)]{akari_irc_cat} Ishihara, D., et al.\ 2010, VizieR Online Data Catalog, 2297, 0 
\bibitem[Ita et al.(2004)]{ita04} Ita Y., et al., 2004, MNRAS, 353, 705 
\bibitem[Ita et al.(2010)]{ita10} Ita, Y., Matsuura, M., Ishihara, D., et al.\ 2010, \aap, 514, A2 
\bibitem[Jackson, Ivezi{\'c}, \& Knapp(2002)]{jackson02} Jackson T., Ivezi{\'c} {\v Z}., Knapp G.~R., 2002, MNRAS, 337, 749 
\bibitem[Joanes \& Gill(1998)]{joanes98} Joanes D.~N., Gill C.~A., 1998, Journal of the Royal Statistical Society (Series D): The Statistician, 47, 183
\bibitem[Jolion, Meer, \& Bataouche(1991)]{jolion91} Jolion J.~M., Meer P., Bataouche S., 1991, IEEE Trans. on Pattern Analysis and Machine Intelligence, 13, 791
\bibitem[Kaiser et al.(2010)]{kaiser10} Kaiser, N., Burgett, W., Chambers, K., et al.\ 2010, \procspie, 7733, 77330E-1
\bibitem[Kelley \& Shaw(2007)]{kelley07} Kelley N., Shaw J.~S.~S., 2007, JSARA, 1, 13 
\bibitem[Kinemuchi et al.(2006)]{kinemuchi06} Kinemuchi K., Smith H.~A., Wo{\'z}niak P.~R., McKay T.~A., 2006, AJ, 132, 1202 
\bibitem[Kinnunen \& Skiff(2000)]{kinnunen00} Kinnunen T., Skiff B.~A., 2000, IBVS, 4865, 1 
\bibitem[Kiss \& Bedding(2003)]{kiss03} Kiss L.~L., Bedding T.~R., 2003, MNRAS, 343, L79 
\bibitem[Kiss et al.(2007)]{kiss07} Kiss L.~L., Derekas A., Szab{\'o} G.~M., Bedding T.~R., Szabados L., 2007, MNRAS, 375, 1338 
\bibitem[Kleinmann(1992)]{kleinmann92} Kleinmann S.~G., 1992, ASPC, 34, 203
\bibitem[Kouzuma \& Yamaoka(2009)]{kouzuma09} Kouzuma S., Yamaoka H., 2009, AJ, 138, 1508 
\bibitem[Kouzuma \& Yamaoka(2010)]{kouzuma10} Kouzuma S., Yamaoka H., 2010, A\&A, 509, A64 
\bibitem[Kwok, Volk, \& Bidelman(1997)]{kwok97} Kwok S., Volk K., Bidelman W.~P., 1997, ApJS, 112, 557 
\bibitem[Liao(2005)]{liao05} Liao T. W., 2005, Pattern Recognition, 38, 1857
\bibitem[Marconi et al.(2006)]{marconi06} Marconi M., Cignoni M., Di Criscienzo M., Ripepi V., Castelli F., Musella I., Ruoppo A., 2006, MNRAS, 371, 1503 
\bibitem[Mauron, Gigoyan, \& Kendall(2007)]{mauron07} Mauron N., Gigoyan K.~S., Kendall T.~R., 2007, A\&A, 475, 843 
\bibitem[Maxted et al.(2009)]{maxted09} Maxted P.~F.~L., G{\"a}nsicke B.~T., Burleigh M.~R., Southworth J., Marsh T.~R., Napiwotzki R., Nelemans G., Wood P.~L., 2009, MNRAS, 400, 2012 
\bibitem[Meyer, Calvet, \& Hillenbrand(1997)]{meyer97} Meyer M.~R., Calvet N., Hillenbrand L.~A., 1997, AJ, 114, 288 
\bibitem[Mickaelian(2008)]{mickaelian08} Mickaelian A.~M., 2008, AJ, 136, 946 
\bibitem[Mochnacki et al.(2002)]{mochnacki02} Mochnacki S.~W., et al., 2002, AJ, 124, 2868
\bibitem[Morrissey et al.(2005)]{morrissey05} Morrissey P., et al., 2005, ApJ, 619, L7 
\bibitem[Morrissey et al.(2007)]{galex} Morrissey P., et al., 2007, ApJS, 173, 682 
\bibitem[Moshir et al.(1990)]{moshir90} Moshir M., et al., 1990, IRAS Faint Source Catalog, version 2.0
\bibitem[Murakami et al.(2007)]{akari} Murakami, H., et al. 2007, PASJ, 59, 369 
\bibitem[Nicholson, Sutherland, \& Sutherland(2005)]{nicholson05} Nicholson M., Sutherland J., Sutherland C., 2005, OEJV, 12, 1 
\bibitem[Nikolaev \& Weinberg(2000)]{nikolaev00} Nikolaev S., Weinberg M.~D., 2000, ApJ, 542, 804 
\bibitem[Olivier, Whitelock, \& Marang(2001)]{oliver01} Olivier E.~A., Whitelock P., Marang F., 2001, MNRAS, 326, 490 
\bibitem[Oyabu et al.(2010)]{akari_cat} Oyabu, S., et al.\ 2010, \procspie, 7731, 77312P-1
\bibitem[Ozawa, Grosso, \& Montmerle(2005)]{ozawa05} Ozawa H., Grosso N., Montmerle T., 2005, A\&A, 429, 963 
\bibitem[Paczy{\'n}ski(2000)]{paczynski00} Paczy{\'n}ski B., 2000, PASP, 112, 1281 
\bibitem[Pojmanski(1997)]{pojmanski97} Pojmanski G., 1997, AcA, 47, 467 
\bibitem[Pojmanski(2002)]{pojmanski02} Pojmanski, G.\ 2002, \actaa, 52, 397 
\bibitem[Pojmanski \& Maciejewski(2005)]{pojmanski05} Pojmanski G., Maciejewski G., 2005, AcA, 55, 97 
\bibitem[Protopapas, Jimenez, \& Alcock(2005)]{protopapas05} Protopapas P., Jimenez R., Alcock C., 2005, MNRAS, 362, 460 
\bibitem[Ramos-Larios et al.(2009)]{ramos-larios09} Ramos-Larios G., Guerrero M.~A., Su{\'a}rez O., Miranda L.~F., G{\'o}mez J.~F., 2009, A\&A, 501, 1207 
\bibitem[Renner et al.(2008)]{renner08} Renner S., Rauer H., Erikson A., Hedelt P., Kabath P., Titz R., Voss H., 2008, A\&A, 492, 617 
\bibitem[Robert(1996)]{robert96} Robert C.~P., 1996, Mixtures of distributions: inference and estimation, in 
Gilks W.~elhalter D.~J., eds, Markov chain Monte Carlo in practice, Chapman \& Hall, London
\bibitem[Schmidt et al.(2009)]{schmidt09} Schmidt E.~G., Hemen B., Rogalla D., Thacker-Lynn L., 2009, AJ, 137, 4598 
\bibitem[Schwarzenberg-Czerny(1996)]{schwarzenberg96} Schwarzenberg-Czerny A., 1996, ApJ, 460, L107 
\bibitem[Seibert et al.(2005)]{seibert05} Seibert M., et al., 2005, ApJ, 619, L23
\bibitem[Sesar et al.(2010)]{sesar10} Sesar B., et al., 2010, ApJ, 708, 717 
\bibitem[Sevenster(2002)]{sevenster02} Sevenster M.~N., 2002, AJ, 123, 2788 
\bibitem[Sharma \& Johnston(2009)]{sharma09} Sharma S., Johnston K.~V., 2009, ApJ, 703, 1061 
\bibitem[Shin \& Byun(2004)]{shin04} Shin M.-S., Byun Y.-I., 2004, JKAS, 37, 79
\bibitem[Shin \& Byun(2007)]{shin07} Shin M.-S., Byun Y.-I., 2007, ASPC, 362, 255
\bibitem[Shin, Sekora, \& Byun(2009)]{paper1} Shin M.-S., Sekora M., Byun Y.-I., 2009, MNRAS, 400, 1897 (Paper I)
\bibitem[Si{\'o}dmiak et al.(2008)]{siodmiak08} Si{\'o}dmiak N., Meixner M., Ueta T., Sugerman B.~E.~K., Van de Steene G.~C., Szczerba R., 2008, ApJ, 677, 382 
\bibitem[Sirko et al.(2004)]{sirko04} Sirko E., et al., 2004, AJ, 127, 899 
\bibitem[Skrutskie et al.(2006)]{2mass} Skrutskie M.~F., et al., 2006, AJ, 131, 1163 
\bibitem[Stephenson(1986)]{stephenson86} Stephenson C.~B., 1986, ApJ, 300, 779 
\bibitem[Sterken \& Jaschek(2005)]{sterken05} Light Curves of Variable Stars, ed. Sterken, C., \& Jaschek, C. (Cambridge, UK: Cambridge University Press)
\bibitem[Stetson(1996)]{stetson96} Stetson P.~B., 1996, PASP, 108, 851
\bibitem[Stoughton et al.(2002)]{stoughton02} Stoughton C., et al., 2002, AJ, 123, 485 
\bibitem[Street et al.(2003)]{street03} Street R.~A., et al., 2003, Scientific Frontiers in Research on Extrasolar Planets, ASP Conference Series, Vol 294, Edited by Drake Deming and Sara Seager. (San Francisco: ASP), pp. 405-408
\bibitem[Szczerba et al.(2007)]{szczerba07} Szczerba R., Si{\'o}dmiak N., Stasi{\'n}ska G., Borkowski J., 2007, A\&A, 469, 799 
\bibitem[Trammell et al.(2007)]{trammell07} Trammell G.~B., Vanden Berk D.~E., Schneider D.~P., Richards G.~T., Hall P.~B., Anderson S.~F., Brinkmann J., 2007, AJ, 133, 1780 
\bibitem[Tsujimoto et al.(2002)]{tsujimoto02} Tsujimoto M., Koyama K., Tsuboi Y., Goto M., Kobayashi N., 2002, ApJ, 566, 974 
\bibitem[Tyson(2002)]{tyson02} Tyson J.~A., 2002, SPIE, 4836, 10 
\bibitem[Usatov(2008)]{usatov08a} Usatov M., 2008, OEJV, 81, 1 
\bibitem[Usatov \& Nosulchik(2008)]{usatov08b} Usatov M., Nosulchik A., 2008, OEJV, 87, 1 
\bibitem[van der Veen \& Habing(1988)]{vanderveen88} van der Veen W.~E.~C.~J., Habing H.~J., 1988, A\&A, 194, 125 
\bibitem[Vendramin, Campello, \& Hruschka(2010)]{vendramin10} Vendramin L., Campello R. J. G. B., Statistical Analysis and Data Mining, 3, 209
\bibitem[V{\'e}ron-Cetty \& V{\'e}ron(2006)]{veron-cetty06} V{\'e}ron-Cetty M.-P., V{\'e}ron P., 2006, A\&A, 455, 773 
\bibitem[Ververidis \& Kotropoulos(2008)]{ververidis08} Ververidis D., Kotropoulos C., 2008, IEEE Trans. on Signal Process., 56, 2797
\bibitem[Vieira et al.(2003)]{vieira03} Vieira S.~L.~A., Corradi W.~J.~B., Alencar S.~H.~P., Mendes L.~T.~S., Torres C.~A.~O., Quast G.~R., Guimar{\~a}es M.~M., da Silva L., 2003, AJ, 126, 2971 
\bibitem[von Neumann(1941)]{vonneumann41} von Neumann J., 1941, The Annals of Mathematical Statistics, 12, 367
\bibitem[Watson(2006)]{watson06} Watson, C.~L.\ 2006, Society for Astronomical Sciences Annual Symposium, 25, 47 
\bibitem[Wegner \& McMahan(1985)]{wegner85} Wegner G., McMahan R.~K., 1985, AJ, 90, 1511 
\bibitem[Welsh et al.(2005)]{welsh05} Welsh, B.~Y., Wheatley, J.~M., Heafield, K., et al.\ 2005, AJ, 130, 825 
\bibitem[Wheatley, Welsh, \& Browne(2008)]{wheatley08} Wheatley J.~M., Welsh B.~Y., Browne S.~E., 2008, AJ, 136, 259 
\bibitem[Wilhelm et al.(1999)]{wilhelm99} Wilhelm R., Beers T.~C., Sommer-Larsen J., Pier J.~R., Layden A.~C., Flynn C., Rossi S., Christensen P.~R., 1999, AJ, 117, 2329 
\bibitem[Williams(1941)]{williams41} Williams J.~D., 1941, The Annals of Mathematical Statistics, 12, 239
\bibitem[Williams et al.(2008)]{williams08} Williams R., Hanisch R., Szalay A., Plante R., 2008, Simple Cone Search Specification Version 1.03, \url{http://www.ivoa.net/Documents/REC/DAL/ConeSearch-20080222.html}
\bibitem[Wils, Lloyd, \& Bernhard(2006)]{wils06} Wils P., Lloyd C., Bernhard K., 2006, MNRAS, 368, 1757 
\bibitem[Wozniak(2000)]{wozniak00} Wozniak P.~R., 2000, AcA, 50, 421 
\bibitem[Wo{\'z}niak et al.(2004a)]{wozniak04a} Wo{\'z}niak P.~R., et al., 2004, AJ, 127, 2436 
\bibitem[Wo{\'z}niak et al.(2004b)]{wozniak04b} Wo{\'z}niak P.~R., Williams S.~J., Vestrand W.~T., Gupta V., 2004, AJ, 128, 2965 
\bibitem[Yamamura et al.(2010)]{akari_fis_cat} Yamamura, I., Makiuti, S., Ikeda, N., Fukuda, Y., Oyabu, S., Koga, T., \& White, G.~J.\ 2010, VizieR Online Data Catalog, 2298, 0 
\bibitem[Yu et al.(2006)]{yu06} Yu J., Amores J., Sebe N., Tian Q., 2006, A New Study on Distance Metrics as Similarity Measurement, in Proceedings of ICME 2006, 533
\bibitem[Yuan \& Akerlof(2008)]{yuan08} Yuan F., Akerlof C.~W., 2008, ApJ, 677, 808 
\bibitem[Zoccali et al.(2003)]{zoccali03} Zoccali M., et al., 2003, A\&A, 399, 931 
\bibitem[Zuckerman(1987)]{zuckerman87} Zuckerman B., 1987, Lecture Notes in Physics, 291, 351 
\end{thebibliography}
\end{document}